\documentclass[10pt,journal,compsoc]{IEEEtran}
\ifCLASSOPTIONcompsoc
  \usepackage[nocompress]{cite}
\else
  \usepackage{cite}
\fi
\usepackage{caption}
\usepackage{subcaption}
\usepackage{graphicx}
\usepackage{tikz}
\usepackage{amssymb}
\usepackage{amsmath}

\DeclareMathOperator*{\argmin}{arg\,min}
\usepackage{amsthm}
\usepackage{amsfonts}
\usepackage{mathrsfs}
\usepackage{txfonts}
\usepackage{stfloats}
\usepackage{cases}
\usepackage{mathtools}
\usepackage{caption}
\usepackage{enumerate}	
\usepackage{enumitem}
\usepackage{longtable}
\usepackage{multirow}
\usepackage{algorithm}
\usepackage{algpseudocode}
\usepackage{enumitem}
\usepackage{mathtools}
\usepackage{listings}
\usepackage{wrapfig}
\usepackage{color}                                               \usepackage{array}                                            \usepackage{longtable}                                        \usepackage{calc}                                             \usepackage{multirow}                                         \usepackage{hhline}                                          
\usepackage{ifthen}                                           

\theoremstyle{definition}
\newtheorem{theorem}{Theorem}[section]

\newtheorem{proposition}{Proposition}[section]
\newtheorem{corollary}[theorem]{Corollary}

\newtheorem{lemma}[theorem]{Lemma}

\ifCLASSINFOpdf
 
\else
  
\fi

\usepackage{url}
\begin{document}
\bibliographystyle{IEEEtran}

\title{Scalable Consensus Protocols for \\PoW based Blockchain and blockDAG}

\author{B~Swaroopa~Reddy,~\IEEEmembership{Member,~IEEE}
        and~G V V~Sharma,~\IEEEmembership{Member,~IEEE}
 
\IEEEcompsocitemizethanks{\IEEEcompsocthanksitem B. Swaroopa Reddy and G V V Sharma are with the Department
of Electrical Engineering, Indian Institute of Technology Hyderabad,  Hyderabad,
Telangana, 502285, India.\protect\\
E-mail: ee17resch11004@iith.ac.in and gadepall@iith.ac.in
}
%\thanks{Manuscript received July 10, 2020; revised xxxxxx XX, YYYY.}
}

% The paper headers
%\markboth{IEEE Transactions on Services Computing,~Vol.~XX, No.~YY, July~2020}%

\IEEEtitleabstractindextext{
\begin{abstract}

This paper proposes two models for scaling the transaction throughput in Proof-of-Work (PoW) based blockchain networks. A mathematical model has been derived for optimal transaction throughput for PoW based longest chain rule blockchain in the first approach. In this approach, the blockchain Peer-to-Peer (P2P) network is considered as Erd$\ddot{o}$s - R$\acute{e}$nyi random network topology. However, this approach is limited by the block creation rate, and the results suggest that the rate beyond an optimal point can result in unfairness in the system. The second approach is a new consensus protocol proposed by considering the ledger as a Directed Acyclic Graph (DAG) called blockDAG instead of a chain of blocks. This framework follows a two-step strategy that makes the system robust enough to handle the double-spend attacks. The first step involves developing an unsupervised learning graph clustering algorithm for separating the blocks created by an attacker. In the second step, after eliminating the attacker's blocks, the remaining blocks are arranged in topological order by honest clients, making the blockDAG ledger system suitable for smart contract applications found in Internet of Things (IoT) services. The simulation results demonstrate a significant improvement in the transaction throughput compared to bitcoin and the network's fairness.

\end{abstract}

}

% make the title area
\maketitle
\IEEEdisplaynontitleabstractindextext
\IEEEpeerreviewmaketitle

\IEEEraisesectionheading{\section{Introduction}\label{sec:introduction}}
\IEEEPARstart{S}{atoshi Nakamoto} introduced blockchain as a P2P network for cryptocurrencies like bitcoin \cite{bitcoin}.  It has also found applications in smart contract-based decentralized applications (DApp) like medical records \cite{medical} and IoT applications \cite{IoT}. The bitcoin protocol in \cite{bitcoin} consists of a chain of blocks in which each block includes a reference to the previous block in the form of a block hash. The miners in the network collect multiple transactions from the clients and create a block by solving a computationally challenging problem called PoW. While creating a new block, this protocol restricts the miner  from referring to the longest chain's tip to maintain consensus among all the network nodes. Each client in the network validates the transactions through local copies of the blocks present with them. The difficulty of the PoW task is adaptively set so that a block is created approximately once every 10 minutes in the entire
network. 

As per the measurement study conducted in \cite{info}, 10 minutes block interval was very high compared to the P2P network's  delay diameter ($D$). Because of this overestimation of delay diameter, bitcoin's consensus rule  \cite{bitcoin} has severe scalability limitations in terms of the number of transactions processed per second (TPS). Bitcoin's network processes around $300000$  transactions in a day \cite{tps} which limits the TPS to $3-4$. Another major issue with the bitcoin protocol is the double-spend attack \cite{Rosenfeld}, where  an attacker creates a secret chain with a block having a transaction that is a replacement for the original payment transaction to a merchant. An attacker publishes her secret chain with high probability after receiving the product from the merchant. To counter the double-spend attack the merchant should wait till the block with his transaction issued by the attacker achieves the required number of confirmations \cite{Rosenfeld} before delivering his product to the attacker.

In this paper, we propose two models for improving the transaction throughput in PoW based blockchain networks without compromising the fairness of the blockchain system.

We propose a mathematical model for optimizing the TPS for the longest chain rule bitcoin protocol in the first model. We framed an optimization problem with the main chain growth rate as a cost function and double-spend attack characterized by attacker's hash rate $q$, delay diameter $D$ as constraints. The $D$ is modeled by considering the blockchain as an Erd$\ddot{o}$s-R$\acute{e}$nyi random network topology \cite{E-R}. The main chain growth rate is derived as a concave function \cite{Boyd} of block-creation rate ($\lambda$).  The TPS is obtained as a function of $\lambda$, block size ($b$), and $D$.  Our model ensures an increase in $\lambda$, resulting in improved throughput. Through simulation results, it is shown that this is achieved without disturbing the balance between the hash rate and rewards of miners. 

However, accelerating $\lambda$ beyond the optimal value proposed in this model causes more blocks in the network. This results in some honest nodes  not having all the blocks created in the network and not extending the longest chain. This makes the system unfair in terms of the balance between hash rates and the rewards of miners. On the other hand, with a high probability, the attacker who does not follow the bitcoin protocol will increase her chain and gain from the double-spend attack \cite{Rosenfeld}.  
To overcome this effect in \textit{the longest chain rule} protocol, we propose a second model named Unsupervised Learning based consensus protocol for blockDAG (UL-blockDAG), a consensus protocol for DAG structure of the ledger instead of a chain of blocks. 

In the second model, while creating a new block, the miner includes reference to all its predecessor blocks called tips/leaf blocks (observed locally in its DAG) in its block header. This results in a DAG structure similar to the blockDAG structure in SPECTRE \cite{spectre} and PHANTOM \cite{phantom}. In this protocol, each client constructs a blockDAG structure and simultaneously
follows the two-step strategy to achieve the consensus. In the first step, each client (node) applies a Graph clustering algorithm \cite{cluster} for two clusters and separates the blocks with less inter-connectivity in the DAG. The intuition for applying the Graph clustering algorithm is to counter the double-spend attack. The attacker does not publish the blocks with double-spend transactions immediately after the creation. Instead, the attacker maintains its ledger secretly until the main ledger reaches the required number of confirmations. So, the attacker's secret chain does not have well-connected blocks, and the client can easily separate those blocks by applying the unsupervised learning-based Graph
clustering algorithms in \cite{cluster}. Finally, the client left with the blocks with well-connectivity, which are called honest blocks. In the second step, the ordering algorithm arranges the honest blocks in topological order, which makes this consensus mechanism suitable for smart contract applications, especially in the Internet of Things (IoT) services due to fast confirmation times. The simulation results show that each honest client counter-attack the attacker's double-spend strategy by executing the unsupervised learning-based clustering algorithm. The results also show that the drastic improvement in the TPS without compromising the fairness of the network.

The rest of the paper is organized as follows. In section 2, we present the related work. Section 3 describes the System model and parameters. Section 4 describes the preliminaries for optimal throughput in the longest chain rule and unsupervised learning based-clustering algorithm. In section 5, we present the optimal longest chain rule. In section 6, we present an unsupervised learning based consensus mechanism for blockDAG. Section 7 discuss simulation results. Section 8 concludes the paper and
give future directions for research.
\section{Related Work}
A GHOST (Greedy Heaviest Observed Sub-tree) rule was proposed in  \cite{ghost},  where, instead of the longest chain consensus rule, the path of the subtree with the combined hardest PoW is chosen. A variant of the GHOST protocol was used in the Ethereum blockchain \cite{ethereum}. However, the TPS for this protocol is still relatively less \cite{ether} and susceptible to attacks described in \cite{on_trees}.

A SPECTRE protocol is proposed in \cite{spectre}, which builds with the concept of the blockDAG structure of the ledger. This protocol describes the ordering of the blocks in the DAG based on the pairwise voting procedure. The pairwise ordering relation is not transitive, and it does not always give the complete linear ordering of the blocks due to the Condorcet paradox \cite{condorcet}. Smart contracts are the general computations on a particular task between two or more parties in the network. In order to achieve the consensus on the computations,  smart contracts need linear ordering of the blocks/transactions. 
Even though the TPS is much higher than \cite{bitcoin} and \cite{ghost}, SPECTRE is not suitable for smart contract applications.

A PHANTOM protocol is proposed in \cite{phantom} based on similar lines of SPECTRE, where, instead of pairwise ordering in SPECTRE, maximum k-cluster subDAG algorithm followed by ordering algorithm were proposed in PHANTOM. While, PHANTOM gives high transaction throughput and total ordering the blocks. However, there is a trade-off between \textit{the anti-cone size} parameter \textit{k} and confirmation times and also susceptible to liveness attack described in \cite{conflux}.

A blockchain or DAG protocol based on delayed rewards  proposed in \cite{delayed_rewards} to address security against double-spend attacks.  The protocol creates a staking mechanism similar to proof-of-stake by delaying the rewards and also punishes the attacker using \textit{fraud-proofs} by slashing or eliminating the miner's future rewards. 

A permissioned blockchain, Hyperledger Fabric, was proposed in \cite{fabric} with Practical Byzantine Fault Tolerant (PBFT) distributed consensus protocol for running distributed applications. PBFT achieves high transaction throughput but is limited to only a few nodes due to communication overhead.

A cryptocurrency for the Internet of Things called IOTA,  which is scalable, lightweight, and Quantum secure, was proposed in \cite{iota} based on the DAG structure of transactions. However, the initial IOTA project \cite{iota} does not design with a smart contract layer. The IOTA team has built the smart contracts on layer-2 \cite{iota_smart} as off-chain smart contracts.
\section{System model and Notations}
The parameters used in our frameworks are listed in TABLE \ref{table:symbols}.

In this work, we stick to Bitcoin's model \cite{bitcoin} in every aspect -- transactions, blocks, PoW, information propagation in the P2P network \cite{info}. The only difference between the optimal longest chain rule and unsupervised learning based consensus protocol is the ledger structure. The first one is similar to the bitcoin
protocol.  In the second framework, instead of a reference
to a single block, the newly created block references more than one block in its header and generates the ledger's DAG structure.

We refer to the  models  in \cite{ghost},\cite{E-R} for  delay diameter $D$, where the bitcoin P2P network  is considered as an Erd$\ddot{o}$s - R$\acute{e}$nyi random network topology. The model specified in \cite{spectre} and \cite{phantom} is used for blockDAG structure of the ledger, where the miners and other nodes follow the mining protocol - 
\begin{enumerate}[]
\item When a miner/node creates/receives a block, broadcast it to all neighboring peers in $V$. More formally, if $c_1, c_2 \in V$, then $G_t^{c_1} \subseteq G_{t+D}^{c_2}$.
\item while creating a new block, the miner includes references to all the leaf-blocks/tips (observed in its local copy of the DAG ($G_t^c$)) in its block header.
\end{enumerate}
As per the mining protocol, every new block $\mathbf{B}$ reaches the entire network in time $D$. While creating a new block at time $t$, the honest miner refers to all the published blocks before time $t-D$,  i.e., $tips \in G_{t-D}^{c}$.  If $\mathbf{B'}$s miner is honest, then it broadcast $\mathbf{B}$ to all its neighboring peers. So, any block $\mathbf{C}$ created after time $t+D$ refer $\mathbf{B}$ directly or indirectly.
Even though the blocks created between $[t-D, t+D]$ are neither referred by $\mathbf{B}$ nor referred $\mathbf{B}$, the $\mathbf{B'}$s miner receives all those blocks by the time $t+2D$.

\begin{table}[!t]
\renewcommand{\arraystretch}{1.3}
\caption{ \\ Parameters used in the proposed the frame works}
\centering
\begin{tabular}{c l}
\hline
\textbf{Symbols}& \textbf{Description} \\
\hline
$n$ & Number of nodes in the network \\
%\hline
$q$ & Fraction of the attacker's hashrate \\
%\hline
$N_t$ & Number of peers connected to each node/client \\
%\hline
$P_e$ & Probability of existence of an edge between two nodes \\
%\hline
$b$ & Block size in \textbf{MB} \\
%\hline
$R$ & Upload bandwidth \\
%\hline
$h$   & Depth of the tree in \cite{delay} \\
%\hline
$T_p$ & Maximum Prpogation delay between a pair of nodes \\
%\hline
$p_c$   & Computational power with node $c$\\
%\hline
$\lambda$ & Block creation rate \\
%\hline
$\beta$  & Main chain growth rate \\
%\hline
$G$ & Block DAG \\
%\hline 
$G_U$ & Undirected version of $G$ \\ 
%\hline
$C$ & Set of the blocks in $G/G_U$ \\
%\hline
$E$ & References to blocks in $G/G_U$ \\
%\hline
$C_1$ & Blocks created by honest nodes $(C_1 \subseteq C)$ \\
%\hline
$C_2$ & Blocks created by attacker (malicious) nodes $(C_2 \subset C)$ \\
%\hline
$\textbf{A}_\textbf{d}$ & Adjacency matrix of blockDAG \\
%\hline
$\textbf{A}$ & Symmetrized Adjacency matrix for $G_U$ \\
%\hline
$\textbf{D}$ & Degree matrix \\
%\hline 
$d_i$ & $i^{th}$ diagonal entry in $\textbf{D}$ \\
%\hline
$\textbf{L}$ & Laplacian matrix \\
%\hline
$\textbf{x}$ & Cluster indicator vector for $C_1$ and $C_2$ \\
%\hline
$\lambda_i$ & $i^{th}$ eigen value of $\textbf{L}$ \\
%\hline
$D$ & End-to-end delay in the network \\
%\hline
$c$ & Client/Node in the network \\
%\hline
$V$ & Set of the clients/nodes in the blockchain network \\
$G_t^c$ & Locally observed blockDAG at client/nodes $c \in V$ at time $t$ \\
$\epsilon$ & Probability of successful double-spend attack \\
$k$ & Number of confirmations \\
$N$ & Height of the blockchain or blockDAG \\
$m$ & Number of mining nodes in the network \\
\hline
\end{tabular}
\label{table:symbols}
\end{table}
\section{Preliminaries}
\subsection{P2P network preliminaries}
\begin{lemma}
Consider the blockchain P2P network as an Erd$\ddot{o}$s-R$\acute{e}$nyi random network as in \cite{E-R}, the worst-case end-to-end block propagation time in the network as per \cite{delay} is 
\end{lemma}
\begin{align}
\label{eq:D}
D = h \left(T_p + \frac{b}{R} N_t\right),
\end{align}
where the degree of the node\footnote{The popular blockchain networks (like bitcoin) were
implemented with a fixed maximum number of peers ($N_t = 8$).} ($N_t$) is derived from a binomial distribution as
\begin{align}
N_t &= (n-1)P_e
\end{align}
and
\begin{align}
h &= \lceil{\log_{N_t}\big(n(N_t - 1)+1 \big)}\rceil   
\end{align}
where, $\lceil . \rceil$ is a ceiling function\footnote{For all $x \in \mathbb{R}$, $\lceil x \rceil$ is the  least integer number greater than or equal to the given $x$.}.

\subsection{Spectral graph theory preliminaries}
The following definitions of spectral graph theory are used in our work.
Let $G = (C, E)$ be a directed graph with vertex set $C$ and edge set $E$. We assumed that the graph is unweighted.\\ 
\textbf{Definition 1. (Adjacency matrix).} The adjacency matrix  $\textbf{A}_\textbf{d}$ of a directed graph $G$ is an $\mid C\mid \times \mid C\mid$ matrix such that
\begin{equation}
(\textbf{A}_\textbf{d})_{ij} = 
\begin{cases}
    1,& \text{if } (i,j)\hspace{0.05cm} \epsilon \hspace{0.05cm} E\\
    0,              & \text{otherwise}
\end{cases}
\end{equation}
where, $\mid C\mid$ is the total number of vertices\footnote{In this framework, Blocks are the vertices of the graph $G$/$G_U$.} of the graph $G_U$.\\
\textbf{Definition 2. (Symmetrization).} The symmetric adjacency matrix  $\textbf{A}$ from $\textbf{A}_\textbf{d}$ without changing the number of edges \cite{symmetrization} is obtained as follows 
\begin{equation}
\mathbf{A = A_d + A_d^T},
\end{equation}
where, $\mathbf{A_d^T}$ is the transpose of the matrix $\mathbf{A_d}$,1 and $\textbf{A}$ represents the adjacency matrix of undirected graph ($G_U$) for the original directed graph $G$. \\
\textbf{Definition 3. (Degree Matrix).} The degree matrix $\textbf{D}$ of the graph (\textbf{$G_U$}) is obtained from $\mathbf{A}$ as follows 
\begin{equation}
\mathbf{D} = \textbf{diag}\{d_1,d_2,\dots,d_{\mid C \mid}\}
\end{equation}
where,
\begin{equation}
d_i = \sum_{j=1}^{\mid C \mid} \textbf{A}_{ij}
\end{equation}
\textbf{Definition 4. (Laplacian Matrix).} The graph Laplacian matrix $\textbf{L}$ is defined as
\begin{equation}
\mathbf{L = D - A}
\end{equation}
Th important properties of $\textbf{L}$ are defined in \cite{cluster}. \\
\textbf{Definition 5. {Rayleigh Ratio}.} The main tool in the optimization problem for graph clustering is Rayleigh ratio \cite{R-R} defined as
\begin{equation}
R(L) = \mathbf{\frac{x^T L x}{x^T x}} \\
\end{equation}
and
\begin{equation}
\lambda_{min} < R(L) < \lambda_{max}
\end{equation}
Where,
\begin{itemize}
\item $\mathbf{x} \in \mathbb{R}^{\mid C \mid}$ is an orthonormal eigen vector of $\textbf{L}$ also used as cluster indicator vector in graph clustering problem.
\item $\lambda_{min}$ and $\lambda_{max}$ are minimum and maximum eigen values of $\textbf{L}$.
\end{itemize}
\section{Optimal Throughput in Longest chain rule protocol}
This section describes the first framework -- a mathematical model for optimizing the TPS in the longest chain rule blockchain protocol. 
The delay diameter $D$ is modeled by assuming the blockchain P2P network as an Erd$\ddot{o}$s - R$\acute{e}$nyi random network topology. The lower-bound on the  main chain growth rate $\beta$ of the longest chain rule network is obtained as a function of $\lambda$ and $D$. 

The TPS is defined as a function of $\beta$, $b$ and $K$. The optimal $\lambda$ for scaling the TPS is obtained by considering the $D$ and double-spend attack characterized by $q$ as constraints for the optimization problem. The constraint to counter the double-spend attack is defined as  inequality between the main chain growth rate $\beta$ and the blocks created by an attacker with a proportion of hash rate $q$ of the total hash rate of the network.

\begin{lemma}
\label{lemma:beta}
For the longest chain rule blockchain network with a block creation rate $\lambda$ and delay diameter $D$,
the lower bound on the main chain growth rate $\beta$ for a small $\delta$ ($\delta << 1$) is 
\begin{equation}
\beta \geq \dfrac{\lambda}{1 + \lambda D - \dfrac{3+\delta}{\sqrt{N}}  }
\label{eq:beta_lim}
\end{equation}
\end{lemma}
\proof See Appendix A.
\begin{corollary}
The number of blocks\footnote{The blockchain's height and the number of blocks are the same in the longest chain rule consensus protocol.} created in the network increase with time. For $N\to\infty$
\begin{align}
\label{eq:beta_inf}
\beta \geq \dfrac{\lambda}{1+\lambda D}.
\end{align}
\end{corollary}
\begin{theorem}
\label{them:tps}
In a blockchain network with throughput $TPS(\lambda,b)$ and delay diameter $D$, 
the optimal block creation rate is
\begin{align}
\label{eq:opt_lam_N}
\lambda &= \frac{1}{D} 
\end{align}
and the optimal transaction throughput is given by 
\begin{align}
TPS(\lambda,b)  = 
\frac{b K}{2 h \left(T_p + \frac{b}{R} N_t\right)}
\label{eq:opt_tps} 
\end{align}
\end{theorem}
\proof
Our goal is to maximize the number of transactions per second $TPS(\lambda,b)$ with very low probability of successful double-spend attack. 

So, the optimization problem can be framed as 
\begin{align}
\label{eq:opt}
\max_{\lambda} \quad TPS(\lambda,b) 
\\
\text{s.t} \quad  q < \frac{\beta}{\lambda}
\label{eq:constaint1}
\\
\frac{1}{\lambda} \geq D
\label{eq:constaint2}
\end{align} 
where, $q\lambda$ is the attacker's chain growth rate and $\frac{1}{\lambda}$ is the block creation interval. 
\begin{align}
\because TPS(\lambda,b)  = \beta  b  K, 
\end{align}
substituting for $\beta$ from \eqref{eq:beta_lim} in \eqref{eq:opt}, the optimal $\lambda$ is obtained by solving
\begin{align}
\label{eq:opt_lam}
\max_{\lambda} \quad \frac{\lambda}{1 - \dfrac{3+\delta}{\sqrt{N}} + \lambda D } 
\\
\text{s.t} \quad  q \left(1 - \frac{3+\delta}{\sqrt{N}} + \lambda D \right) < 1
\\
\lambda D \leq 1
\end{align} 
The solution to the optimization problem is provided in Appendix B.
\section{The Unsupervised Learning-based blockDAG (UL-blockDAG) consensus protocol}
In this section, we discuss the operation of the proposed blockDAG consensus protocol. The protocol consists of the following two steps-
\begin{itemize}
\item Separating the blocks (with less inter-connectivity) created by an attacker from the well-connected blocks in the blockDAG using the spectral graph theory-based Unsupervised Learning algorithm.
\item A topological ordering of the blocks based on the directed edges in DAG structure.
\end{itemize}
\subsection{Spectral graph clustering for DAG}
The graph clustering is defined in two ways by maximizing the intra-cluster edge connections and minimizing the inter-cluster edge connections. Consider an undirected and unweighted graph $G_U$ of a blockDAG $G=(C,E)$. The input data to find the similarities between the intra-cluster vertices is the adjacency matrix $A$. The objective is to find the cluster indicator vector $\mathbf{x} = (x_1, x_2, \dots, x_{|C|})$, which maps vertices of graph $G_U$ to real numbers so that $(x_i - x_j)^2$ should be small (maximal connected vertices stay as close together as possible).

The optimization problem for graph clustering based on spectral graph theory \cite{opt_prob}, \cite{Laplacian} is 
\begin{align}
\label{eq:opt_cluster}
\argmin_{\mathbf{x} \in \mathbb{R}^{\mid C \mid} } \quad & \mathbf{x^T L x} \\
\text{s.t} \quad & \mathbf{x^T 1} = 0 \label{eq:c1} \\
&
\mathbf{x^T x} = 1 \label{eq:c2}
\end{align}

The solution to the above optimization problem with the objective function $\argmin_{\mathbf{x} \in \mathbb{R}^{\mid C \mid} } \mathbf{x^T L x}$ gives a trivial solution \cite{Laplacian} $\mathbf{x}=\mathbf{1}$  for an eigenvalue $\lambda_{min} = 0$. This causes uncertainty in clustering the graph's vertices by mapping all vertices of $G$ with a constant value $x_i = 1$, for all $i$. The constraint in \eqref{eq:c1} eliminates  this uncertainty. The \textit{sign} of the elements of the eigenvector $\mathbf{x}$ decides the clustering process and there is no constraint on the magnitude of $\mathbf{x}$. Therefore, the normalization \eqref{eq:c2} is imposed on \eqref{eq:opt_cluster}. Thus, by the Rayleigh-Ritz theorem \cite{R-R}, the solution to the above optimization problem is the eigenvector ($\mathbf{x}$) corresponding to the second smallest eigenvalue ($\lambda_2$) of $\mathbf{L}$ (the smallest non-zero eigenvalue). 

Finally, the \textit{sign} of the real value $x_i$ (for all $i$) maps the vertices to the two clusters $C_1$ and $C_2$ of graph $G$. This entire graph clustering procedure is shown in Algorithm \ref{Algorithm1}. The intuition for using Algorithm \ref{Algorithm1} for separating the attacker blocks is explained through the following example.

\begin{algorithm}[!t]
  \caption{Spectral graph clustering for DAG}\label{Algorithm1}
  \textbf{Input:} G - DAG \\
  \textbf{output:} $C_1, C_2$
  \begin{algorithmic}[1]
    \Procedure{FIND-CLUSTERS}{$G$}  
      \State Construct $\textbf{A}_\textbf{d}$ for $G$
		\State Compute $\textbf{A} = \textbf{A}_\textbf{d} + \textbf{A}_\textbf{d}^\textbf{T}$
		\State Compute the Laplacian matrix $\textbf{L} = \textbf{D}-\textbf{A}$
		\State Compute $\textbf{x}$ \Comment eigen vector corresponding to 
		       $2^{nd}$ \hspace*{3.3cm} smallest eigen value of $\textbf{L}$
		 \For {$x_i \in \mathbf{x}$}       
		 \If{$x_i >= 0$}
				\State $C_1.append$($vertex \in C$ mapped to $x_i$)
			\Else 
				\State $C_2.append$($vertex \in C$ mapped to $x_i$)
		    \EndIf
		    \EndFor
      \State \textbf{return} $C_1$, $C_2$ 
    \EndProcedure
  \end{algorithmic}
\end{algorithm}
   
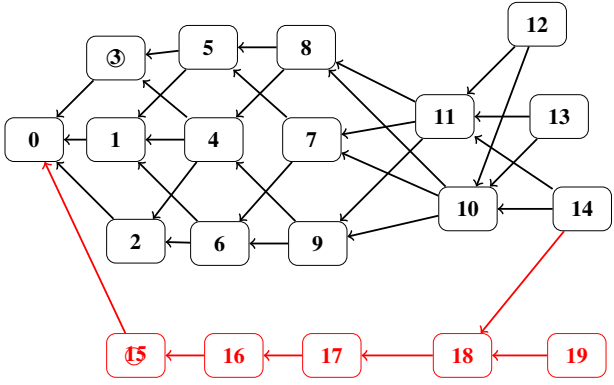
\begin{figure}[h]
\centering
\resizebox{\columnwidth}{!}{
\tikzstyle{block} = [square, draw,
    text width=7em, text centered, minimum height=4em]
\tikzstyle{sum} = [draw, circle, node distance=3cm]
\tikzstyle{input} = [coordinate]
\tikzstyle{output} = [coordinate]
\tikzstyle{pinstyle} = [pin edge={to-,thin,black}]
\tikzstyle{line} = [draw, -latex']

\tikzset{
   bigbigbox/.style = {minimum width=2.5cm, rectangle},
   bigbox/.style = {draw,rectangle},
   box/.style = {minimum width=2.7cm, rounded corners,rectangle, fill=blue!20},
   square/.style = {minimum width=15mm,minimum height=15mm}
   }

\begin{tikzpicture}[auto, node distance=2cm]
\node [rectangle, draw,text width=2em, text centered,rounded corners, minimum height=2em,name=genesis] {$\textbf{0}$} ;

\node [rectangle, draw,text width=2em, text centered,rounded corners, minimum height=2em, below right of=genesis,node distance=2.2cm] (2) {$\textbf{2}$};
     
\node [rectangle, draw,text width=2em, text centered,rounded corners, minimum height=2em, right of=genesis,node  distance=1.25cm] (1) {$\textbf{1}$};

\node [rectangle, draw,text width=2em, text centered,rounded corners, minimum height=2em, above of=1,node distance=1.25cm] (3) {$\textcircled{\textbf{3}}$};
\draw [->,thick] (1) -- node {} (genesis);
\draw [->,thick] (2) -- node {} (genesis);
\draw [->,thick] (3) -- node {} (genesis);
\node [rectangle, draw,text width=2em, text centered,rounded corners, minimum height=2em, below right of=1,node distance=2.25cm] (6) {$\textbf{6}$};

\node [rectangle, draw,text width=2em, text centered,rounded corners, minimum height=2em, right of=1,node distance=1.5cm] (4) {$\textbf{4}$};
\node [rectangle, draw,text width=2em, text centered,rounded corners, minimum height=2em, above right of=1,node distance=2cm] (5) {$\textbf{5}$};
\node [rectangle, draw,text width=2em, text centered,rounded corners, minimum height=2em, right of=4,node distance=1.5cm] (7) {$\textbf{7}$};
\draw [->,thick] (4) -- node {} (2);
\draw [->,thick] (5) -- node {} (1);
\draw [->,thick] (6) -- node {} (1);
\draw [->,thick] (4) -- node {} (1);
\draw [->,thick] (6) -- node {} (2);
\draw [->,thick] (5) -- node {} (3);
\draw [->,thick] (4) -- node {} (3);
%
%\draw [->,thick] (5) -- node {} (4);
%\draw [->,thick] (Y) -- node {} (4);
%\draw [->,thick] (Y) -- node {} (3);
%
\node [rectangle, draw,text width=2em, text centered,rounded corners, minimum height=2em, above right of=4,node distance=2 cm] (8) {$\textbf{8}$};
\node [rectangle, draw,text width=2em, text centered,rounded corners, minimum height=2em, right of=6,node distance=1.5cm] (9) {$\textbf{9}$};
\node [rectangle, draw,text width=2em, text centered,rounded corners, minimum height=2em, below right of=8,node distance=3.5cm] (10) {$\textbf{10}$};
\node [rectangle, draw,text width=2em, text centered,rounded corners, minimum height=2em, above right of=9,node distance=2.75cm] (11) {$\textbf{11}$};
\node [rectangle, draw,text width=2em, text centered,rounded corners, minimum height=2em, right of=10,node distance=1.75cm] (14) {$\textbf{14}$};
\node [rectangle, draw,text width=2em, text centered,rounded corners, minimum height=2em, right of=11,node distance=1.75cm] (13) {$\textbf{13}$};
\node [rectangle, draw,text width=2em, text centered,rounded corners, minimum height=2em, above right of=11,node distance=2cm] (12) {$\textbf{12}$};
\draw [->,thick] (8) -- node {} (4);
\draw [->,thick] (8) -- node {} (5);
\draw [->,thick] (7) -- node {} (6);
\draw [->,thick] (7) -- node {} (5);
\draw [->,thick] (10) -- node {} (7);
\draw [->,thick] (10) -- node {} (9);
\draw [->,thick] (9) -- node {} (6);
\draw [->,thick] (9) -- node {} (4);
\draw [->,thick] (10) -- node {} (8);
\draw [->,thick] (11) -- node {} (8);
\draw [->,thick] (11) -- node {} (7);
\draw [->,thick] (11) -- node {} (9);

\draw [->,thick] (12) -- node {} (10);
\draw [->,thick] (12) -- node {} (11);
\draw [->,thick] (13) -- node {} (11);
\draw [->,thick] (13) -- node {} (10);
\draw [->,thick] (14) -- node {} (10);
\draw [->,thick] (14) -- node {} (11);

\node [rectangle, draw,text width=2em, text centered,rounded corners, minimum height=2em, below of=2,color=red,node distance=1.75cm] (15) {$\textcircled{\textbf{15}}$};
\node [rectangle, draw,text width=2em, text centered,rounded corners, minimum height=2em, right of=15,color=red,node distance=1.5cm] (16) {$\textbf{16}$};
\node [rectangle, draw,text width=2em, text centered,rounded corners, minimum height=2em,right of=16,color=red,node distance=1.5cm] (17) {$\textbf{17}$};
\node [rectangle, draw,text width=2em, text centered,rounded corners, minimum height=2em, right of=17,color=red,node distance=2cm] (18) {$\textbf{18}$};
\node [rectangle, draw,text width=2em, text centered,rounded corners, minimum height=2em, right of=18,color=red,node distance=1.75cm] (19) {$\textbf{19}$};
\draw [->,thick,color=red] (15) -- node {} (genesis);
\draw [->,thick,color=red] (16) -- node {} (15);
\draw [->,thick,color=red] (17) -- node {} (16);
\draw [->,thick,color=red] (18) -- node {} (17);
\draw [->,thick,color=red] (14) -- node {} (18);
\draw [->,thick,color=red] (19) -- node {} (18);

\end{tikzpicture}
}
\caption{An example of a blockDAG attacker model.  Block $\textbf{15}$ created by an attacker has a conflicting transaction with a transaction in block $\textbf{3}$ shown by rounded text. The blocks from $\mathbf{0-14}$ are generated by honest miners and an attacker keeps the blocks from $\textbf{15}-\textbf{17}$ in secret until $\textbf{3}$ attained sufficient confirmations (In this case $3$ confirmations) and broadcast all the blocks created by the attacker. Observe that there are no references to attacker blocks $\textbf{15}-\textbf{17}$ from the blocks created by honest nodes indicates blocks $\textbf{15}-\textbf{17}$ created in secret, but both the honest block ($\textbf{14}$) and the attacker block ($\textbf{19}$) has a reference to block $\textbf{18}$ indicate the attacker broadcasted the blocks after the creation of block $\mathbf{18}$.}
\label{fig:attacker_model}
\end{figure}

An attacker model of blockDAG structure is shown in Fig. \ref{fig:attacker_model}. The red-colored blocks show the blocks created by an attacker in secret to attempt the double-spend attack.  The attacker creates a
block with a conflicting transaction\footnote{The transactions which claim the same sources of cryptocurrency are called conflicting transactions. For example, in a bitcoin network, a node that acts as an attacker attempts a double-spend attack by creating two conflicting transactions with the same Unspent Transaction Output (UTXO) \cite{bitcoin}.} for an honest block's original transaction and creates a chain of blocks in secret. When the block with an original transaction attained the required number of confirmations, the attacker broadcast the secret chain.  

\begin{figure}[!t]
\centerline{\includegraphics[width=\columnwidth]{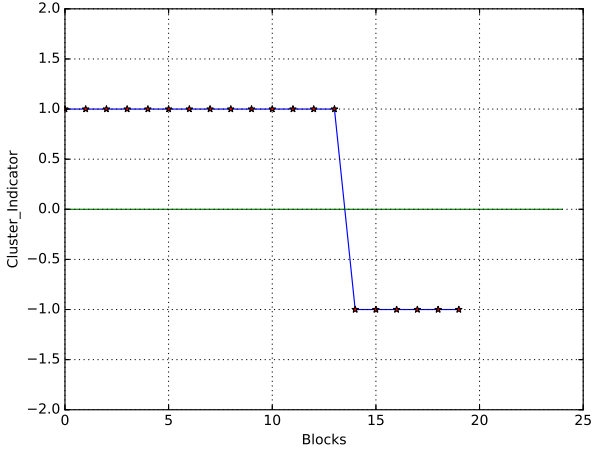}}
\caption{Grouping the blocks in blockDAG shown in Fig. \ref{fig:attacker_model} into two clusters. The blocks above the green line  are added to $C_1$ and blocks below the green line are added to $C_2$.}
\label{fig:eigen_vector}
\end{figure}

Fig. \ref{fig:eigen_vector} shows the output of Algorithm \ref{Algorithm1} for the blockDAG $G$ has shown in Fig. \ref{fig:attacker_model}. The positive and negative elements of the eigenvector $\mathbf{x}$  for the blockDAG $G$ in Fig. \ref{fig:attacker_model} are mapped to discrete values in the set $\{1,-1\}$ which represents the clusters $C_1$ and $C_2$. The blocks from $\textbf{0}-\textbf{13}$ added to $C_1$ and the attacker blocks $\textbf{15}-\textbf{19}$ along with block $\textbf{14}$ that reference  $\textbf{18}$ (an attacker block) are added to $C_2$.
  
The order of the matrix $A_d$ is $|C| \times |C|$. In Algorithm \ref{Algorithm1}, construction of $A_d$, computing $A$ and $L$ will take $\mathcal{O}(|C|^2)$ operations and computing $\mathbf{x}$  will take  $\mathcal{O}(|C|^3)$ operations. Finally, sorting the eigenvalues will be computed in $\mathcal{O}(|C| \log |C|)$ operations and finding $C_1$ and $C_2$ will take $\mathcal{O}(|C|)$ operations. So, Algorithm \ref{Algorithm1} for finding the clusters $C_1$ and $C_2$ from the blockDAG $G$ will be computed in $\mathcal{O}(|C|^3)$ operations. As the block creation is a Poisson random process (See Lemma \ref{lemma:beta_dag}), the size of vertices $|C|$ of $G$ depends on $\lambda$.

%%%%
\subsection{The client protocol}
A client/node is either a miner having a high computational power to solve the PoW problem or a simple node with no mining power. The client protocol is a two-step consensus protocol described through the following two algorithms.
\subsubsection{Separating the attacker blocks and honest blocks} 
\begin{figure}[!t]
\begin{subfigure}{.45\columnwidth}
\resizebox{\textwidth}{!}{
  %\documentclass{article}
%\usepackage{tikz}
%\usetikzlibrary{shapes,arrows}
%\usetikzlibrary{arrows.meta,calc,fit,positioning}
%\newcommand*\circled[1]{\tikz[baseline=(char.base)]{
%            \node[shape=circle,draw,inner sep=0.5pt] (char) {#1};}}
%\begin{document}
\tikzstyle{block} = [square, draw,
    text width=7em, text centered, minimum height=4em]
\tikzstyle{sum} = [draw, circle, node distance=3cm]
\tikzstyle{input} = [coordinate]
\tikzstyle{output} = [coordinate]
\tikzstyle{pinstyle} = [pin edge={to-,thin,black}]
\tikzstyle{line} = [draw, -latex']

\tikzset{
   bigbigbox/.style = {minimum width=2.5cm, rectangle},
   bigbox/.style = {draw,rectangle},
   box/.style = {minimum width=2.7cm, rounded corners,rectangle, fill=blue!20},
   square/.style = {minimum width=15mm,minimum height=15mm}
   }

\begin{tikzpicture}[auto, node distance=2cm]
\node [rectangle, draw,text width=2em, text centered,rounded corners, minimum height=2em,name=genesis] {$\textbf{0}$} ;

\node [rectangle, draw,text width=2em, text centered,rounded corners, minimum height=2em, below right of=genesis,node distance=1.5cm] (2) {$\textbf{2}$};
     
\node [rectangle, draw,text width=2em, text centered,rounded corners, minimum height=2em, right of=genesis,node  distance=1.25cm] (1) {$\textbf{1}$};

\node [rectangle, draw,text width=2em, text centered,rounded corners, minimum height=2em, above of=1,node distance=1.25cm] (3) {$\textbf{3}$};
\draw [->,thick] (1) -- node {} (genesis);
\draw [->,thick] (2) -- node {} (genesis);
\draw [->,thick] (3) -- node {} (genesis);

\node [rectangle, draw,text width=2em, text centered,rounded corners, minimum height=2em, color = red,right of=1,node distance=1.75cm] (H) {$\textbf{H}$};
\draw [->,thick,color=red] (H) -- node {} (1);
\draw [->,thick,color=red] (H) -- node {} (2);
\draw [->,thick,color=red] (H) -- node {} (3);
\end{tikzpicture}
%\end{document} 
  }
\caption{At height=1}
\label{fig:h1}
\end{subfigure} 
\hfill
\begin{subfigure}{.45\columnwidth}
\resizebox{\textwidth}{!}{
  %\documentclass{article}
%\usepackage{tikz}
%\usetikzlibrary{shapes,arrows}
%\usetikzlibrary{arrows.meta,calc,fit,positioning}
%\newcommand*\circled[1]{\tikz[baseline=(char.base)]{
%            \node[shape=circle,draw,inner sep=0.5pt] (char) {#1};}}
%\begin{document}
\tikzstyle{block} = [square, draw,
    text width=7em, text centered, minimum height=4em]
\tikzstyle{sum} = [draw, circle, node distance=3cm]
\tikzstyle{input} = [coordinate]
\tikzstyle{output} = [coordinate]
\tikzstyle{pinstyle} = [pin edge={to-,thin,black}]
\tikzstyle{line} = [draw, -latex']

\tikzset{
   bigbigbox/.style = {minimum width=2.5cm, rectangle},
   bigbox/.style = {draw,rectangle},
   box/.style = {minimum width=2.7cm, rounded corners,rectangle, fill=blue!20},
   square/.style = {minimum width=15mm,minimum height=15mm}
   }

\begin{tikzpicture}[auto, node distance=2cm]
\node [rectangle, draw,text width=2em, text centered,rounded corners, minimum height=2em,name=genesis] {$\textbf{0}$} ;

\node [rectangle, draw,text width=2em, text centered,rounded corners, minimum height=2em, below right of=genesis,node distance=1.5cm] (2) {$\textbf{2}$};
     
\node [rectangle, draw,text width=2em, text centered,rounded corners, minimum height=2em, right of=genesis,node  distance=1.25cm] (1) {$\textbf{1}$};

\node [rectangle, draw,text width=2em, text centered,rounded corners, minimum height=2em, above of=1,node distance=1.25cm] (3) {$\textbf{3}$};
\draw [->,thick] (1) -- node {} (genesis);
\draw [->,thick] (2) -- node {} (genesis);
\draw [->,thick] (3) -- node {} (genesis);
\node [rectangle, draw,text width=2em, text centered,rounded corners, minimum height=2em, below right of=1,node distance=2cm] (6) {$\textbf{6}$};

\node [rectangle, draw,text width=2em, text centered,rounded corners, minimum height=2em, right of=1,node distance=1.5cm] (4) {$\textbf{4}$};
\node [rectangle, draw,text width=2em, text centered,rounded corners, minimum height=2em, above right of=1,node distance=2cm] (5) {$\textbf{5}$};
\node [rectangle, draw,text width=2em, text centered,rounded corners, minimum height=2em, right of=4,color=red,node distance=1.5cm] (H) {$\textbf{H}$};
\draw [->,thick] (4) -- node {} (2);
\draw [->,thick] (5) -- node {} (1);
\draw [->,thick] (6) -- node {} (1);
\draw [->,thick] (4) -- node {} (1);
\draw [->,thick] (6) -- node {} (2);
\draw [->,thick] (5) -- node {} (3);
\draw [->,thick] (4) -- node {} (3);
\draw [->,thick,color=red] (H) -- node {} (4);
\draw [->,thick,color=red] (H) -- node {} (5);
\draw [->,thick,color=red] (H) -- node {} (6);
\end{tikzpicture}
%\end{document}
  }
\caption{At height=2}
\label{fig:h2}
\end{subfigure}
\\
\begin{subfigure}{.45\columnwidth}
\resizebox{\textwidth}{!}{
  %\documentclass{article}
%\usepackage{tikz}
%\usetikzlibrary{shapes,arrows}
%\usetikzlibrary{arrows.meta,calc,fit,positioning}
%\newcommand*\circled[1]{\tikz[baseline=(char.base)]{
%            \node[shape=circle,draw,inner sep=0.5pt] (char) {#1};}}
%\begin{document}
\tikzstyle{block} = [square, draw,
    text width=7em, text centered, minimum height=4em]
\tikzstyle{sum} = [draw, circle, node distance=3cm]
\tikzstyle{input} = [coordinate]
\tikzstyle{output} = [coordinate]
\tikzstyle{pinstyle} = [pin edge={to-,thin,black}]
\tikzstyle{line} = [draw, -latex']

\tikzset{
   bigbigbox/.style = {minimum width=2.5cm, rectangle},
   bigbox/.style = {draw,rectangle},
   box/.style = {minimum width=2.7cm, rounded corners,rectangle, fill=blue!20},
   square/.style = {minimum width=15mm,minimum height=15mm}
   }

\begin{tikzpicture}[auto, node distance=2cm]
\node [rectangle, draw,text width=2em, text centered,rounded corners, minimum height=2em,name=genesis] {$\textbf{0}$} ;

\node [rectangle, draw,text width=2em, text centered,rounded corners, minimum height=2em, below right of=genesis,node distance=1.5cm] (2) {$\textbf{2}$};
     
\node [rectangle, draw,text width=2em, text centered,rounded corners, minimum height=2em, right of=genesis,node  distance=1.25cm] (1) {$\textbf{1}$};

\node [rectangle, draw,text width=2em, text centered,rounded corners, minimum height=2em, above of=1,node distance=1.25cm] (3) {$\textbf{3}$};
\draw [->,thick] (1) -- node {} (genesis);
\draw [->,thick] (2) -- node {} (genesis);
\draw [->,thick] (3) -- node {} (genesis);
\node [rectangle, draw,text width=2em, text centered,rounded corners, minimum height=2em, below right of=1,node distance=2cm] (6) {$\textbf{6}$};

\node [rectangle, draw,text width=2em, text centered,rounded corners, minimum height=2em, right of=1,node distance=1.5cm] (4) {$\textbf{4}$};
\node [rectangle, draw,text width=2em, text centered,rounded corners, minimum height=2em, above right of=1,node distance=2cm] (5) {$\textbf{5}$};
\node [rectangle, draw,text width=2em, text centered,rounded corners, minimum height=2em, right of=4,node distance=1.5cm] (7) {$\textbf{7}$};
\draw [->,thick] (4) -- node {} (2);
\draw [->,thick] (5) -- node {} (1);
\draw [->,thick] (6) -- node {} (1);
\draw [->,thick] (4) -- node {} (1);
\draw [->,thick] (6) -- node {} (2);
\draw [->,thick] (5) -- node {} (3);
\draw [->,thick] (4) -- node {} (3);
%
%\draw [->,thick] (5) -- node {} (4);
%\draw [->,thick] (Y) -- node {} (4);
%\draw [->,thick] (Y) -- node {} (3);
%
\node [rectangle, draw,text width=2em, text centered,rounded corners, minimum height=2em, above right of=4,node distance=2 cm] (8) {$\textbf{8}$};
\node [rectangle, draw,text width=2em, text centered,rounded corners, minimum height=2em, right of=6,node distance=1.5cm] (9) {$\textbf{9}$};

\draw [->,thick] (8) -- node {} (4);
\draw [->,thick] (8) -- node {} (5);
\draw [->,thick] (7) -- node {} (6);
\draw [->,thick] (7) -- node {} (5);
\draw [->,thick] (9) -- node {} (6);
\draw [->,thick] (9) -- node {} (4);

\node [rectangle, draw,text width=2em, text centered,rounded corners, minimum height=2em, right of=7,color=red,node distance=1.5cm] (H) {$\textbf{H}$};
\draw [->,thick,color=red] (H) -- node {} (9);
\draw [->,thick,color=red] (H) -- node {} (8);
\draw [->,thick,color=red] (H) -- node {} (7);
\end{tikzpicture}
%\end{document}
  }
\caption{At height=3}
\label{fig:h3}
\end{subfigure}
\hfill
\begin{subfigure}{.45\columnwidth}
\resizebox{\textwidth}{!}{
  %\documentclass{article}
%\usepackage{tikz}
%\usetikzlibrary{shapes,arrows}
%\usetikzlibrary{arrows.meta,calc,fit,positioning}
%\newcommand*\circled[1]{\tikz[baseline=(char.base)]{
%            \node[shape=circle,draw,inner sep=0.5pt] (char) {#1};}}
%\begin{document}
\tikzstyle{block} = [square, draw,
    text width=7em, text centered, minimum height=4em]
\tikzstyle{sum} = [draw, circle, node distance=3cm]
\tikzstyle{input} = [coordinate]
\tikzstyle{output} = [coordinate]
\tikzstyle{pinstyle} = [pin edge={to-,thin,black}]
\tikzstyle{line} = [draw, -latex']

\tikzset{
   bigbigbox/.style = {minimum width=2.5cm, rectangle},
   bigbox/.style = {draw,rectangle},
   box/.style = {minimum width=2.7cm, rounded corners,rectangle, fill=blue!20},
   square/.style = {minimum width=15mm,minimum height=15mm}
   }

\begin{tikzpicture}[auto, node distance=2cm]
\node [rectangle, draw,text width=2em, text centered,rounded corners, minimum height=2em,name=genesis] {$\textbf{0}$} ;

\node [rectangle, draw,text width=2em, text centered,rounded corners, minimum height=2em, below right of=genesis,node distance=1.5cm] (2) {$\textbf{2}$};
     
\node [rectangle, draw,text width=2em, text centered,rounded corners, minimum height=2em, right of=genesis,node  distance=1.25cm] (1) {$\textbf{1}$};

\node [rectangle, draw,text width=2em, text centered,rounded corners, minimum height=2em, above of=1,node distance=1.25cm] (3) {$\textbf{3}$};
\draw [->,thick] (1) -- node {} (genesis);
\draw [->,thick] (2) -- node {} (genesis);
\draw [->,thick] (3) -- node {} (genesis);
\node [rectangle, draw,text width=2em, text centered,rounded corners, minimum height=2em, below right of=1,node distance=2cm] (6) {$\textbf{6}$};

\node [rectangle, draw,text width=2em, text centered,rounded corners, minimum height=2em, right of=1,node distance=1.5cm] (4) {$\textbf{4}$};
\node [rectangle, draw,text width=2em, text centered,rounded corners, minimum height=2em, above right of=1,node distance=2cm] (5) {$\textbf{5}$};
\node [rectangle, draw,text width=2em, text centered,rounded corners, minimum height=2em, right of=4,node distance=1.5cm] (7) {$\textbf{7}$};
\draw [->,thick] (4) -- node {} (2);
\draw [->,thick] (5) -- node {} (1);
\draw [->,thick] (6) -- node {} (1);
\draw [->,thick] (4) -- node {} (1);
\draw [->,thick] (6) -- node {} (2);
\draw [->,thick] (5) -- node {} (3);
\draw [->,thick] (4) -- node {} (3);
%
%\draw [->,thick] (5) -- node {} (4);
%\draw [->,thick] (Y) -- node {} (4);
%\draw [->,thick] (Y) -- node {} (3);
%
\node [rectangle, draw,text width=2em, text centered,rounded corners, minimum height=2em, above right of=4,node distance=2 cm] (8) {$\textbf{8}$};
\node [rectangle, draw,text width=2em, text centered,rounded corners, minimum height=2em, right of=6,node distance=1.5cm] (9) {$\textbf{9}$};
\node [rectangle, draw,text width=2em, text centered,rounded corners, minimum height=2em, below right of=8,node distance=3cm] (10) {$\textbf{10}$};
\node [rectangle, draw,text width=2em, text centered,rounded corners, minimum height=2em, above right of=9,node distance=2.75cm] (11) {$\textbf{11}$};

\draw [->,thick] (8) -- node {} (4);
\draw [->,thick] (8) -- node {} (5);
\draw [->,thick] (7) -- node {} (6);
\draw [->,thick] (7) -- node {} (5);
\draw [->,thick] (10) -- node {} (7);
\draw [->,thick] (10) -- node {} (9);
\draw [->,thick] (9) -- node {} (6);
\draw [->,thick] (9) -- node {} (4);
\draw [->,thick] (10) -- node {} (8);
\draw [->,thick] (11) -- node {} (8);
\draw [->,thick] (11) -- node {} (7);
\draw [->,thick] (11) -- node {} (9);

\node [rectangle, draw,text width=2em, text centered,rounded corners, minimum height=2em, right of=7,color=red,node distance=3.5cm] (H) {$\textbf{H}$};
\draw [->,thick,color=red] (H) -- node {} (11);
\draw [->,thick,color=red] (H) -- node {} (10);
\end{tikzpicture}
%\end{document}
  }
\caption{At height=4}
\label{fig:h4}
\end{subfigure}
\\
\begin{subfigure}{.45\columnwidth}
\resizebox{\textwidth}{!}{
  %\documentclass{article}
%\usepackage{tikz}
%\usetikzlibrary{shapes,arrows}
%\usetikzlibrary{arrows.meta,calc,fit,positioning}
%\newcommand*\circled[1]{\tikz[baseline=(char.base)]{
%            \node[shape=circle,draw,inner sep=0.5pt] (char) {#1};}}
%\begin{document}
\tikzstyle{block} = [square, draw,
    text width=7em, text centered, minimum height=4em]
\tikzstyle{sum} = [draw, circle, node distance=3cm]
\tikzstyle{input} = [coordinate]
\tikzstyle{output} = [coordinate]
\tikzstyle{pinstyle} = [pin edge={to-,thin,black}]
\tikzstyle{line} = [draw, -latex']

\tikzset{
   bigbigbox/.style = {minimum width=2.5cm, rectangle},
   bigbox/.style = {draw,rectangle},
   box/.style = {minimum width=2.7cm, rounded corners,rectangle, fill=blue!20},
   square/.style = {minimum width=15mm,minimum height=15mm}
   }

\begin{tikzpicture}[auto, node distance=2cm]
\node [rectangle, draw,text width=2em, text centered,rounded corners, minimum height=2em,name=genesis] {$\textbf{0}$} ;

\node [rectangle, draw,text width=2em, text centered,rounded corners, minimum height=2em, below right of=genesis,node distance=1.5cm] (2) {$\textbf{2}$};
     
\node [rectangle, draw,text width=2em, text centered,rounded corners, minimum height=2em, right of=genesis,node  distance=1.25cm] (1) {$\textbf{1}$};

\node [rectangle, draw,text width=2em, text centered,rounded corners, minimum height=2em, above of=1,node distance=1.25cm] (3) {$\textbf{3}$};
\draw [->,thick] (1) -- node {} (genesis);
\draw [->,thick] (2) -- node {} (genesis);
\draw [->,thick] (3) -- node {} (genesis);
\node [rectangle, draw,text width=2em, text centered,rounded corners, minimum height=2em, below right of=1,node distance=2cm] (6) {$\textbf{6}$};

\node [rectangle, draw,text width=2em, text centered,rounded corners, minimum height=2em, right of=1,node distance=1.5cm] (4) {$\textbf{4}$};
\node [rectangle, draw,text width=2em, text centered,rounded corners, minimum height=2em, above right of=1,node distance=2cm] (5) {$\textbf{5}$};
\node [rectangle, draw,text width=2em, text centered,rounded corners, minimum height=2em, right of=4,node distance=1.5cm] (7) {$\textbf{7}$};
\draw [->,thick] (4) -- node {} (2);
\draw [->,thick] (5) -- node {} (1);
\draw [->,thick] (6) -- node {} (1);
\draw [->,thick] (4) -- node {} (1);
\draw [->,thick] (6) -- node {} (2);
\draw [->,thick] (5) -- node {} (3);
\draw [->,thick] (4) -- node {} (3);
%
%\draw [->,thick] (5) -- node {} (4);
%\draw [->,thick] (Y) -- node {} (4);
%\draw [->,thick] (Y) -- node {} (3);
%
\node [rectangle, draw,text width=2em, text centered,rounded corners, minimum height=2em, above right of=4,node distance=2 cm] (8) {$\textbf{8}$};
\node [rectangle, draw,text width=2em, text centered,rounded corners, minimum height=2em, right of=6,node distance=1.5cm] (9) {$\textbf{9}$};
\node [rectangle, draw,text width=2em, text centered,rounded corners, minimum height=2em, below right of=8,node distance=3cm] (10) {$\textbf{10}$};
\node [rectangle, draw,text width=2em, text centered,rounded corners, minimum height=2em, above right of=9,node distance=2.75cm] (11) {$\textbf{11}$};
\node [rectangle, draw,text width=2em, text centered,rounded corners, minimum height=2em, right of=10,node distance=1.5cm] (14) {$\textbf{14}$};
\node [rectangle, draw,text width=2em, text centered,rounded corners, minimum height=2em, right of=11,node distance=1.75cm] (13) {$\textbf{13}$};
\node [rectangle, draw,text width=2em, text centered,rounded corners, minimum height=2em, above right of=11,node distance=2cm] (12) {$\textbf{12}$};
\draw [->,thick] (8) -- node {} (4);
\draw [->,thick] (8) -- node {} (5);
\draw [->,thick] (7) -- node {} (6);
\draw [->,thick] (7) -- node {} (5);
\draw [->,thick] (10) -- node {} (7);
\draw [->,thick] (10) -- node {} (9);
\draw [->,thick] (9) -- node {} (6);
\draw [->,thick] (9) -- node {} (4);
\draw [->,thick] (10) -- node {} (8);
\draw [->,thick] (11) -- node {} (8);
\draw [->,thick] (11) -- node {} (7);
\draw [->,thick] (11) -- node {} (9);

\draw [->,thick] (12) -- node {} (10);
\draw [->,thick] (12) -- node {} (11);
\draw [->,thick] (13) -- node {} (11);
\draw [->,thick] (13) -- node {} (10);
\draw [->,thick] (14) -- node {} (10);
\draw [->,thick] (14) -- node {} (11);

%
%\node [rectangle, draw,text width=2em, text centered,rounded corners, minimum height=2em, below of=6,node distance=1.25cm] (15) {$\textbf{15}$};
%%
%\node [rectangle, draw,text width=2em, text centered,rounded corners, minimum height=2em, right of=15,node distance=1.25cm] (16) {$\textbf{16}$};
%%
%\node [rectangle, draw,text width=2em, text centered,rounded corners, minimum height=2em,right of=16,node distance=1.25cm] (17) {$\textbf{17}$};
%%
%\node [rectangle, draw,text width=2em, text centered,rounded corners, minimum height=2em, right of=17,node distance=1.25cm] (18) {$\textbf{18}$};
%%
%\node [rectangle, draw,text width=2em, text centered,rounded corners, minimum height=2em, right of=18,node distance=1.5cm] (19) {$\textbf{19}$};
%%
%\draw [->,thick] (15) -- node {} (2);
%\draw [->,thick] (16) -- node {} (15);
%\draw [->,thick] (17) -- node {} (16);
%\draw [->,thick] (18) -- node {} (17);
%\draw [->,thick] (14) -- node {} (18);
%\draw [->,thick] (19) -- node {} (18);

\node [rectangle, draw,text width=2em, text centered,rounded corners, minimum height=2em, right of=13,color=red,node distance=1.75cm] (H) {$\textbf{H}$};
\draw [->,thick,color=red] (H) -- node {} (14);
\draw [->,thick,color=red] (H) -- node {} (13);
\draw [->,thick,color=red] (H) -- node {} (12);
%\draw [->,thick,color=red] (H) -- node {} (19);
\end{tikzpicture}
%\end{document}
  }
\caption{At height=5}
\label{fig:h5}
\end{subfigure}
\hfill
\begin{subfigure}{.45\columnwidth}
\resizebox{\textwidth}{!}{
  %\documentclass{article}
%\usepackage{tikz}
%\usetikzlibrary{shapes,arrows}
%\usetikzlibrary{arrows.meta,calc,fit,positioning}
%\newcommand*\circled[1]{\tikz[baseline=(char.base)]{
%            \node[shape=circle,draw,inner sep=0.5pt] (char) {#1};}}
%\begin{document}
\tikzstyle{block} = [square, draw,
    text width=7em, text centered, minimum height=4em]
\tikzstyle{sum} = [draw, circle, node distance=3cm]
\tikzstyle{input} = [coordinate]
\tikzstyle{output} = [coordinate]
\tikzstyle{pinstyle} = [pin edge={to-,thin,black}]
\tikzstyle{line} = [draw, -latex']

\tikzset{
   bigbigbox/.style = {minimum width=2.5cm, rectangle},
   bigbox/.style = {draw,rectangle},
   box/.style = {minimum width=2.7cm, rounded corners,rectangle, fill=blue!20},
   square/.style = {minimum width=15mm,minimum height=15mm}
   }

\begin{tikzpicture}[auto, node distance=2cm]
\node [rectangle, draw,text width=2em, text centered,rounded corners, minimum height=2em,color=blue,name=genesis] {$\textbf{0}$} ;

\node [rectangle, draw,text width=2em, text centered,rounded corners, minimum height=2em, below right of=genesis,color=blue,node distance=2.2cm] (2) {$\textbf{2}$};
     
\node [rectangle, draw,text width=2em, text centered,rounded corners, minimum height=2em, right of=genesis,color=blue,node  distance=1.25cm] (1) {$\textbf{1}$};

\node [rectangle, draw,text width=2em, text centered,rounded corners, minimum height=2em, above of=1,color=blue,node distance=1.25cm] (3) {$\textcircled{\textbf{3}}$};
\draw [->,thick,color=blue] (1) -- node {} (genesis);
\draw [->,thick,color=blue] (2) -- node {} (genesis);
\draw [->,thick,color=blue] (3) -- node {} (genesis);
\node [rectangle, draw,text width=2em, text centered,rounded corners, minimum height=2em, below right of=1,node distance=2.25cm] (6) {$\textbf{6}$};

\node [rectangle, draw,text width=2em, text centered,rounded corners, minimum height=2em, right of=1,node distance=1.5cm] (4) {$\textbf{4}$};
\node [rectangle, draw,text width=2em, text centered,rounded corners, minimum height=2em, above right of=1,node distance=2cm] (5) {$\textbf{5}$};
\node [rectangle, draw,text width=2em, text centered,rounded corners, minimum height=2em, right of=4,node distance=1.5cm] (7) {$\textbf{7}$};
\draw [->,thick] (4) -- node {} (2);
\draw [->,thick] (5) -- node {} (1);
\draw [->,thick] (6) -- node {} (1);
\draw [->,thick] (4) -- node {} (1);
\draw [->,thick] (6) -- node {} (2);
\draw [->,thick] (5) -- node {} (3);
\draw [->,thick] (4) -- node {} (3);
%
%\draw [->,thick] (5) -- node {} (4);
%\draw [->,thick] (Y) -- node {} (4);
%\draw [->,thick] (Y) -- node {} (3);
%
\node [rectangle, draw,text width=2em, text centered,rounded corners, minimum height=2em, above right of=4,node distance=2 cm] (8) {$\textbf{8}$};
\node [rectangle, draw,text width=2em, text centered,rounded corners, minimum height=2em, right of=6,node distance=1.5cm] (9) {$\textbf{9}$};
\node [rectangle, draw,text width=2em, text centered,rounded corners, minimum height=2em, below right of=8,node distance=3.5cm] (10) {$\textbf{10}$};
\node [rectangle, draw,text width=2em, text centered,rounded corners, minimum height=2em, above right of=9,node distance=2.75cm] (11) {$\textbf{11}$};
\node [rectangle, draw,text width=2em, text centered,rounded corners, minimum height=2em, right of=10,node distance=1.75cm] (14) {$\textbf{14}$};
\node [rectangle, draw,text width=2em, text centered,rounded corners, minimum height=2em, right of=11,node distance=1.75cm] (13) {$\textbf{13}$};
\node [rectangle, draw,text width=2em, text centered,rounded corners, minimum height=2em, above right of=11,node distance=2cm] (12) {$\textbf{12}$};
\draw [->,thick] (8) -- node {} (4);
\draw [->,thick] (8) -- node {} (5);
\draw [->,thick] (7) -- node {} (6);
\draw [->,thick] (7) -- node {} (5);
\draw [->,thick] (10) -- node {} (7);
\draw [->,thick] (10) -- node {} (9);
\draw [->,thick] (9) -- node {} (6);
\draw [->,thick] (9) -- node {} (4);
\draw [->,thick] (10) -- node {} (8);
\draw [->,thick] (11) -- node {} (8);
\draw [->,thick] (11) -- node {} (7);
\draw [->,thick] (11) -- node {} (9);

\draw [->,thick] (12) -- node {} (10);
\draw [->,thick] (12) -- node {} (11);
\draw [->,thick] (13) -- node {} (11);
\draw [->,thick] (13) -- node {} (10);
\draw [->,thick] (14) -- node {} (10);
\draw [->,thick] (14) -- node {} (11);

\node [rectangle, draw,text width=2em, text centered,rounded corners, minimum height=2em, below of=2,color=red,node distance=1.75cm] (15) {$\textcircled{\textbf{15}}$};
\node [rectangle, draw,text width=2em, text centered,rounded corners, minimum height=2em, right of=15,node distance=1.5cm] (16) {$\textbf{16}$};
\node [rectangle, draw,text width=2em, text centered,rounded corners, minimum height=2em,right of=16,node distance=1.5cm] (17) {$\textbf{17}$};
\node [rectangle, draw,text width=2em, text centered,rounded corners, minimum height=2em, right of=17,node distance=2cm] (18) {$\textbf{18}$};
\node [rectangle, draw,text width=2em, text centered,rounded corners, minimum height=2em, right of=18,node distance=1.75cm] (19) {$\textbf{19}$};
\draw [->,thick,color=red] (15) -- node {} (genesis);
\draw [->,thick] (16) -- node {} (15);
\draw [->,thick] (17) -- node {} (16);
\draw [->,thick] (18) -- node {} (17);
%\draw [->,thick] (14) -- node {} (18);
\draw [->,thick] (19) -- node {} (18);

\node [rectangle, draw,text width=2em, text centered,rounded corners, minimum height=2em, right of=13,node distance=1.75cm] (20) {$\textbf{21}$};
\node [rectangle, draw,text width=2em, text centered,rounded corners, minimum height=2em, right of=12,node distance=2cm] (21) {$\textbf{20}$};
\node [rectangle, draw,text width=2em, text centered,rounded corners, minimum height=2em, above right of=19,node distance=2.5cm] (22) {$\textbf{22}$};
\draw [->,thick] (22) -- node {} (13);
\draw [->,thick] (22) -- node {} (14);
\draw [->,thick] (21) -- node {} (12);
\draw [->,thick] (21) -- node {} (14);
\draw [->,thick] (21) -- node {} (13);
\draw [->,thick] (20) -- node {} (14);
\draw [->,thick] (20) -- node {} (12);
\draw [->,thick] (20) -- node {} (13);
\draw [->,thick] (22) -- node {} (19);
\draw [->,thick] (20) -- node {} (19);

\node [rectangle, draw,text width=2em, text centered,rounded corners, minimum height=2em, right of=20,color=red,node distance=1.75cm] (H) {$\textbf{H}$};
\draw [->,thick,color=red] (H) -- node {} (20);
\draw [->,thick,color=red] (H) -- node {} (21);
\draw [->,thick,color=red] (H) -- node {} (22);
%\draw [->,thick,color=red] (H) -- node {} (19);
\end{tikzpicture}
%\end{document}
  }
\caption{At height=6}
\label{fig:h6}
\end{subfigure}
\\
\begin{subfigure}{.45\columnwidth}
\resizebox{\textwidth}{!}{
  %\documentclass{article}
%\usepackage{tikz}
%\usetikzlibrary{shapes,arrows}
%\usetikzlibrary{arrows.meta,calc,fit,positioning}
%\newcommand*\circled[1]{\tikz[baseline=(char.base)]{
%            \node[shape=circle,draw,inner sep=0.5pt] (char) {#1};}}
%\begin{document}
\tikzstyle{block} = [square, draw,
    text width=7em, text centered, minimum height=4em]
\tikzstyle{sum} = [draw, circle, node distance=3cm]
\tikzstyle{input} = [coordinate]
\tikzstyle{output} = [coordinate]
\tikzstyle{pinstyle} = [pin edge={to-,thin,black}]
\tikzstyle{line} = [draw, -latex']

\tikzset{
   bigbigbox/.style = {minimum width=2.5cm, rectangle},
   bigbox/.style = {draw,rectangle},
   box/.style = {minimum width=2.7cm, rounded corners,rectangle, fill=blue!20},
   square/.style = {minimum width=15mm,minimum height=15mm}
   }

\begin{tikzpicture}[auto, node distance=2cm]
\node [rectangle, draw,text width=2em, text centered,rounded corners, minimum height=2em,color=blue,name=genesis] {$\textbf{0}$} ;

\node [rectangle, draw,text width=2em, text centered,rounded corners, minimum height=2em, below right of=genesis,color=blue,node distance=2.2cm] (2) {$\textbf{2}$};
     
\node [rectangle, draw,text width=2em, text centered,rounded corners, minimum height=2em, right of=genesis,color=blue,node  distance=1.25cm] (1) {$\textbf{1}$};

\node [rectangle, draw,text width=2em, text centered,rounded corners, minimum height=2em, above of=1,color=blue,node distance=1.25cm] (3) {$\textcircled{\textbf{3}}$};
\draw [->,thick,color=blue] (1) -- node {} (genesis);
\draw [->,thick,color=blue] (2) -- node {} (genesis);
\draw [->,thick,color=blue] (3) -- node {} (genesis);
\node [rectangle, draw,text width=2em, text centered,rounded corners, minimum height=2em, below right of=1,color=blue,node distance=2.25cm] (6) {$\textbf{6}$};

\node [rectangle, draw,text width=2em, text centered,rounded corners, minimum height=2em, right of=1,color=blue,node distance=1.5cm] (4) {$\textbf{4}$};
\node [rectangle, draw,text width=2em, text centered,rounded corners, minimum height=2em, above right of=1,color=blue,node distance=2cm] (5) {$\textbf{5}$};
\node [rectangle, draw,text width=2em, text centered,rounded corners, minimum height=2em, right of=4,node distance=1.5cm] (7) {$\textbf{7}$};
\draw [->,thick,color=blue] (4) -- node {} (2);
\draw [->,thick,color=blue] (5) -- node {} (1);
\draw [->,thick,color=blue] (6) -- node {} (1);
\draw [->,thick,color=blue] (4) -- node {} (1);
\draw [->,thick,color=blue] (6) -- node {} (2);
\draw [->,thick,color=blue] (5) -- node {} (3);
\draw [->,thick,color=blue] (4) -- node {} (3);
%
%\draw [->,thick] (5) -- node {} (4);
%\draw [->,thick] (Y) -- node {} (4);
%\draw [->,thick] (Y) -- node {} (3);
%
\node [rectangle, draw,text width=2em, text centered,rounded corners, minimum height=2em, above right of=4,node distance=2 cm] (8) {$\textbf{8}$};
\node [rectangle, draw,text width=2em, text centered,rounded corners, minimum height=2em, right of=6,node distance=1.5cm] (9) {$\textbf{9}$};
\node [rectangle, draw,text width=2em, text centered,rounded corners, minimum height=2em, below right of=8,node distance=3.5cm] (10) {$\textbf{10}$};
\node [rectangle, draw,text width=2em, text centered,rounded corners, minimum height=2em, above right of=9,node distance=2.75cm] (11) {$\textbf{11}$};
\node [rectangle, draw,text width=2em, text centered,rounded corners, minimum height=2em, right of=10,node distance=1.75cm] (14) {$\textbf{14}$};
\node [rectangle, draw,text width=2em, text centered,rounded corners, minimum height=2em, right of=11,node distance=1.75cm] (13) {$\textbf{13}$};
\node [rectangle, draw,text width=2em, text centered,rounded corners, minimum height=2em, above right of=11,node distance=2cm] (12) {$\textbf{12}$};
\draw [->,thick] (8) -- node {} (4);
\draw [->,thick] (8) -- node {} (5);
\draw [->,thick] (7) -- node {} (6);
\draw [->,thick] (7) -- node {} (5);
\draw [->,thick] (10) -- node {} (7);
\draw [->,thick] (10) -- node {} (9);
\draw [->,thick] (9) -- node {} (6);
\draw [->,thick] (9) -- node {} (4);
\draw [->,thick] (10) -- node {} (8);
\draw [->,thick] (11) -- node {} (8);
\draw [->,thick] (11) -- node {} (7);
\draw [->,thick] (11) -- node {} (9);

\draw [->,thick] (12) -- node {} (10);
\draw [->,thick] (12) -- node {} (11);
\draw [->,thick] (13) -- node {} (11);
\draw [->,thick] (13) -- node {} (10);
\draw [->,thick] (14) -- node {} (10);
\draw [->,thick] (14) -- node {} (11);

\node [rectangle, draw,text width=2em, text centered,rounded corners, minimum height=2em, below of=2,color=red,node distance=1.75cm] (15) {$\textcircled{\textbf{15}}$};
\node [rectangle, draw,text width=2em, text centered,rounded corners, minimum height=2em,color=red, right of=15,node distance=1.5cm] (16) {$\textbf{16}$};
\node [rectangle, draw,text width=2em, text centered,rounded corners, minimum height=2em,right of=16,node distance=1.5cm] (17) {$\textbf{17}$};
\node [rectangle, draw,text width=2em, text centered,rounded corners, minimum height=2em, right of=17,node distance=2cm] (18) {$\textbf{18}$};
\node [rectangle, draw,text width=2em, text centered,rounded corners, minimum height=2em, right of=18,node distance=1.75cm] (19) {$\textbf{19}$};
\draw [->,thick,color=red] (15) -- node {} (genesis);
\draw [->,thick,color=red] (16) -- node {} (15);
\draw [->,thick] (17) -- node {} (16);
\draw [->,thick] (18) -- node {} (17);
%\draw [->,thick] (14) -- node {} (18);
\draw [->,thick] (19) -- node {} (18);

\node [rectangle, draw,text width=2em, text centered,rounded corners, minimum height=2em, right of=13,node distance=1.75cm] (20) {$\textbf{21}$};
\node [rectangle, draw,text width=2em, text centered,rounded corners, minimum height=2em, right of=12,node distance=2cm] (21) {$\textbf{20}$};
\node [rectangle, draw,text width=2em, text centered,rounded corners, minimum height=2em, above right of=19,node distance=2.5cm] (22) {$\textbf{22}$};
\draw [->,thick] (22) -- node {} (13);
\draw [->,thick] (22) -- node {} (14);
\draw [->,thick] (21) -- node {} (12);
\draw [->,thick] (21) -- node {} (14);
\draw [->,thick] (21) -- node {} (13);
\draw [->,thick] (20) -- node {} (14);
\draw [->,thick] (20) -- node {} (12);
\draw [->,thick] (20) -- node {} (13);
\draw [->,thick] (22) -- node {} (19);
\draw [->,thick] (20) -- node {} (19);

\node [rectangle, draw,text width=2em, text centered,rounded corners, minimum height=2em, below right of=21,node distance=2cm] (23) {$\textbf{23}$};
\node [rectangle, draw,text width=2em, text centered,rounded corners, minimum height=2em, below right of=20,node distance=2.5cm] (24) {$\textbf{24}$};
\node [rectangle, draw,text width=2em, text centered,rounded corners, minimum height=2em, above right of=20,node distance=2cm] (25) {$\textbf{25}$};
\draw [->,thick] (25) -- node {} (20);
\draw [->,thick] (25) -- node {} (21);
\draw [->,thick] (23) -- node {} (20);
\draw [->,thick] (23) -- node {} (21);
\draw [->,thick] (23) -- node {} (22);
\draw [->,thick] (24) -- node {} (20);
\draw [->,thick] (24) -- node {} (22);

\node [rectangle, draw,text width=2em, text centered,rounded corners, minimum height=2em, right of=23,color=red,node distance=1.75cm] (H) {$\textbf{H}$};
\draw [->,thick,color=red] (H) -- node {} (23);
\draw [->,thick,color=red] (H) -- node {} (24);
\draw [->,thick,color=red] (H) -- node {} (25);
\end{tikzpicture}
%\end{document}
  }
\caption{At height=7}
\label{fig:h7}
\end{subfigure}
\caption{An example to show the operation of Algorithm 2.  As time progresses, clustering and confirmation of blocks to $C_1$ (blue) and $C_2$ (red).}
\label{fig:algo_2}
\end{figure}
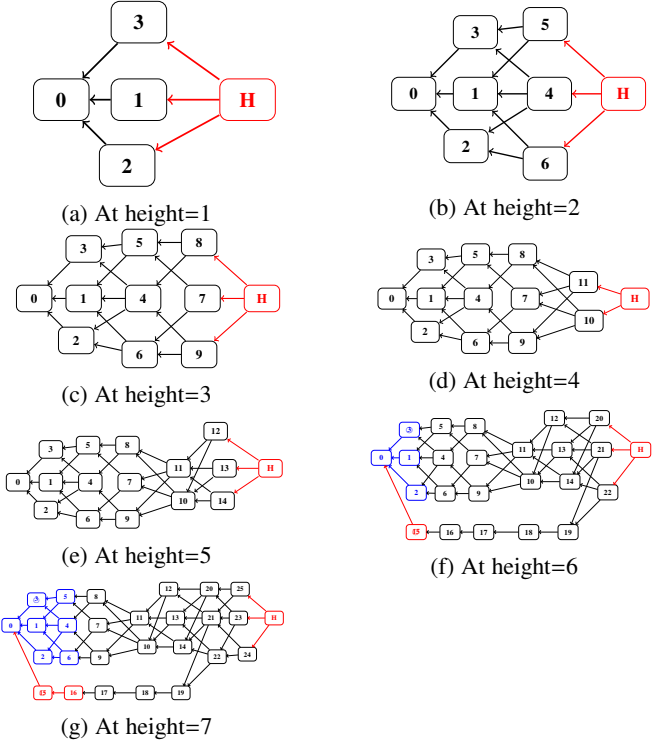

For all $c \in V$, the input $G_t^c$ to Algorithm 2 at each time instant $t$ has shown in Fig. \ref{fig:algo_2}. For simplicity, at each time instant, a hypothetical block $H$ added to $G_t^c$. 
\begin{algorithm}[!t]
  \caption{Finding the list of confirmed blocks}\label{Algorithm2}
  \textbf{Input:} $G_t^c$, $k$, $N$ - Height at which decision on confirmation \\ \hspace{1cm} to be made \\
	\textbf{output:} $\textbf{blueList}$ - A set of confirmed blocks
  \begin{algorithmic}[1]
  	\Procedure{FIND-LIST}{$G^c_t,k,N$} 
         \State $G \gets G^c_t \backslash $ \{All left most blocks till $height = N-1$\}
         \State $C_1, C_2 \gets {FIND-CLUSTERS}(G)$ 
		   \For{all $B \in C_1 \cap$ Blocks at height $N$}
		       \State \textbf{blueList}.add(B)
		   \EndFor
		   \State \textbf{return} $\textbf{blueList}$ 
    \EndProcedure
  \end{algorithmic}
\end{algorithm}
Algorithm \ref{Algorithm2} operates as follows -
\begin{itemize}
\item Given a blockDAG $G_t^c$ observed locally at each client $c$ at a particular time $t$, the algorithm finds the clusters $C_1$ and $C_2$ each time recursively at each block height  after adding new blocks to $G_t^c$.
\item Whenever blocks at a particular height ($N-1$) attain the required number of confirmations, the client excludes those blocks from the input $G_t^c$ to construct $G$. 
\item To decide the confirmation of blocks at height $N$, the client needs to reach the height $N+k+1$ in its local blockDAG $G_t^c$.
Suppose a client observed the current block height $N$ in its local blockDAG $G_t^c$. 
\item To assign the blocks at height $N$ to clusters $C_1$ and $C_2$, the client considers the blocks in heights [$N$, $N+k+1$].
So the symmetric matrix $\mathbf{L}$ is the order of the total number of blocks in the heights [$N$, $N+k+1$] present in $G_t^c$.
\item The confirmed blocks added to the $\textbf{blueList}$ for arranging them into topological order as described in Algorithm \ref{Algorithm3}.
\end{itemize}
Let $C$ is the set of blocks in the graph $G$ of line $2$ in Algorithm \ref{Algorithm2}. Asymptotically, Algorithm \ref{Algorithm2} computes $blueList$ in $\mathcal{O}(|C|^3)$ operations.
 
\begin{figure}[!t]
  \begin{subfigure}[b]{0.45\columnwidth}
    \includegraphics[width=\linewidth]{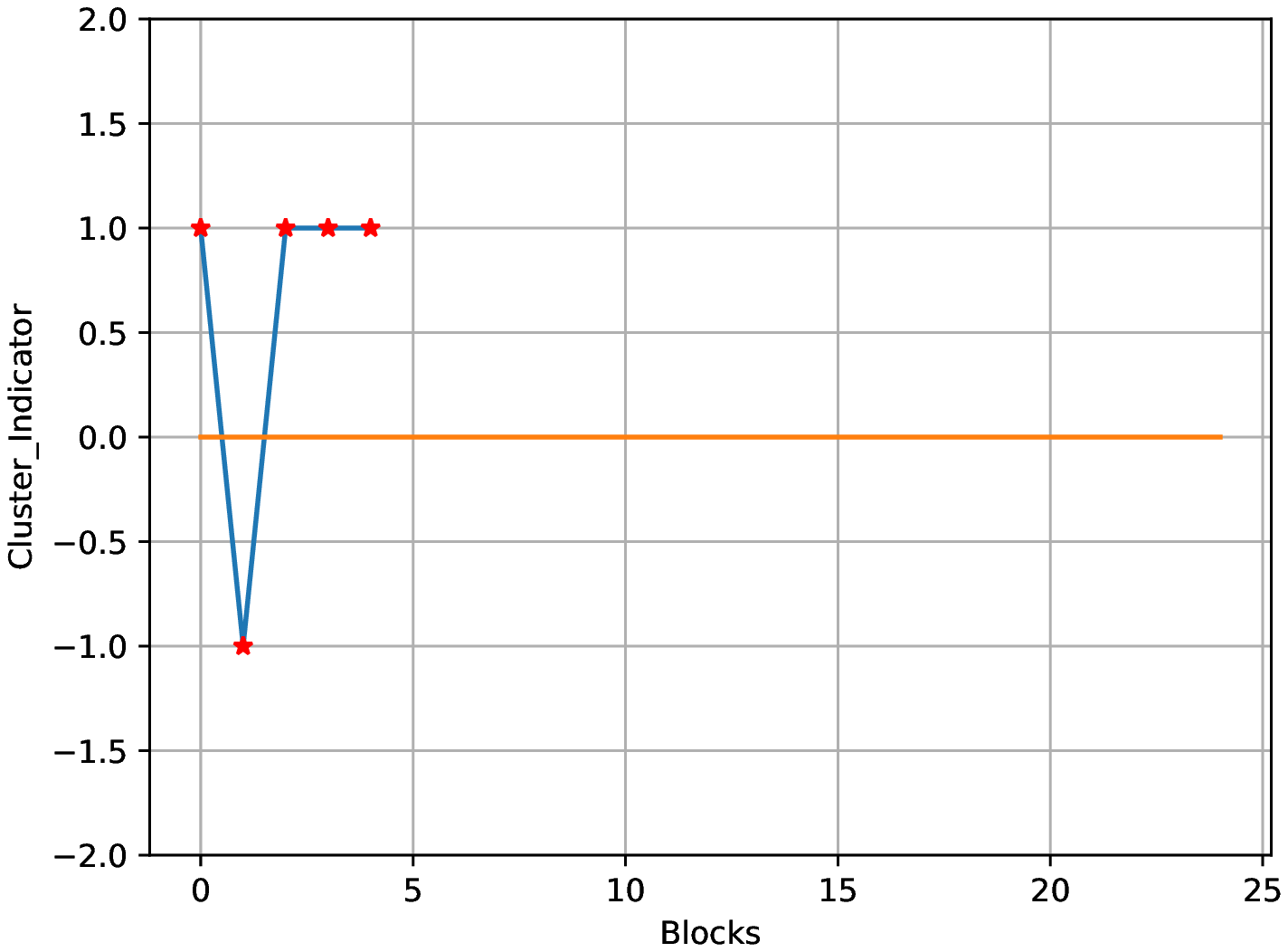}
    \caption{At height = 1}
    \label{fig:h_1}
  \end{subfigure}
  \hfill %%
  \begin{subfigure}[b]{0.45\columnwidth}
    \includegraphics[width=\linewidth]{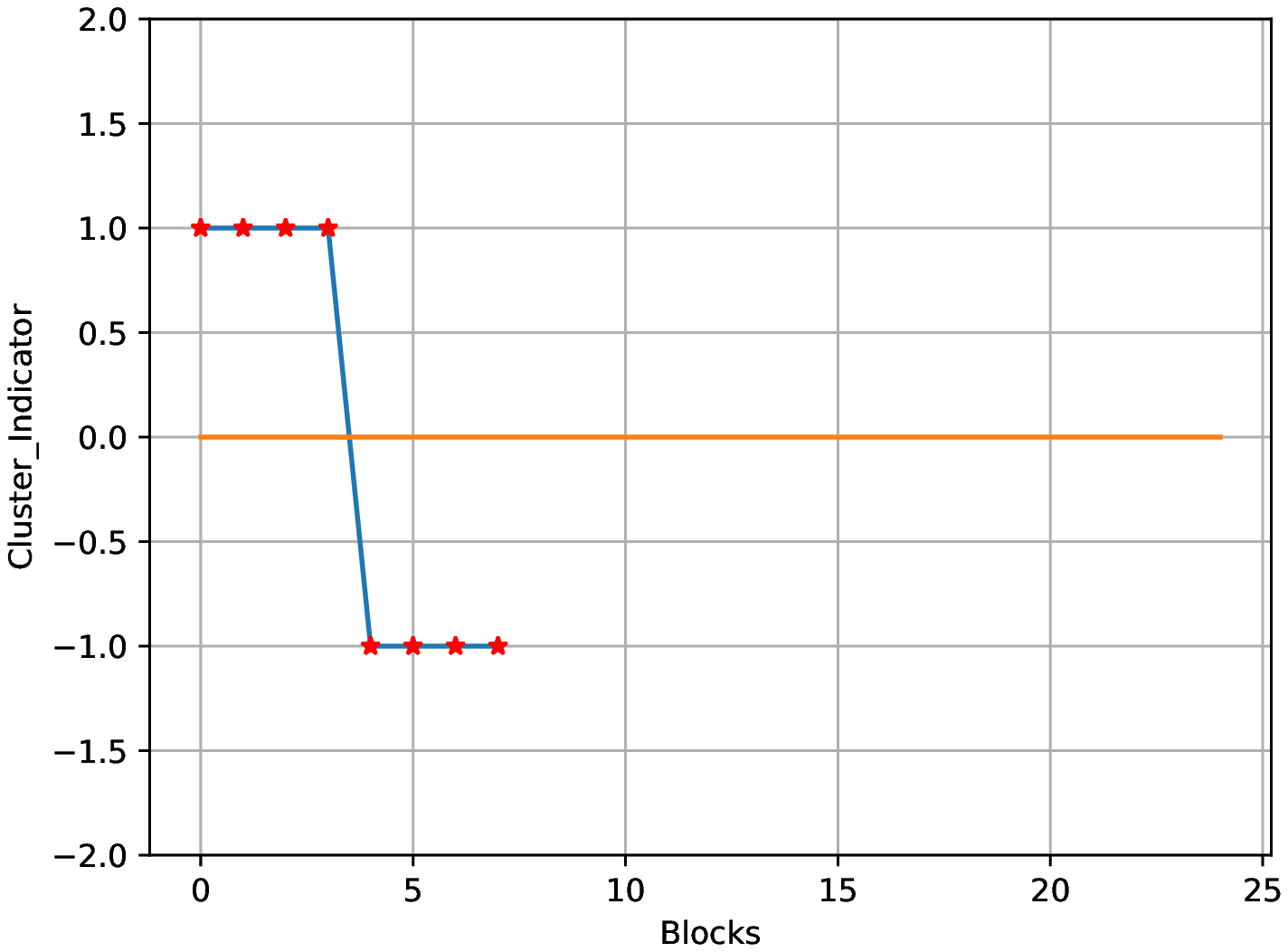}
    \caption{At height = 2}
    \label{fig:h_2}
  \end{subfigure}
  \\
  %\hfill
  \begin{subfigure}[b]{0.45\columnwidth}
    \includegraphics[width=\linewidth]{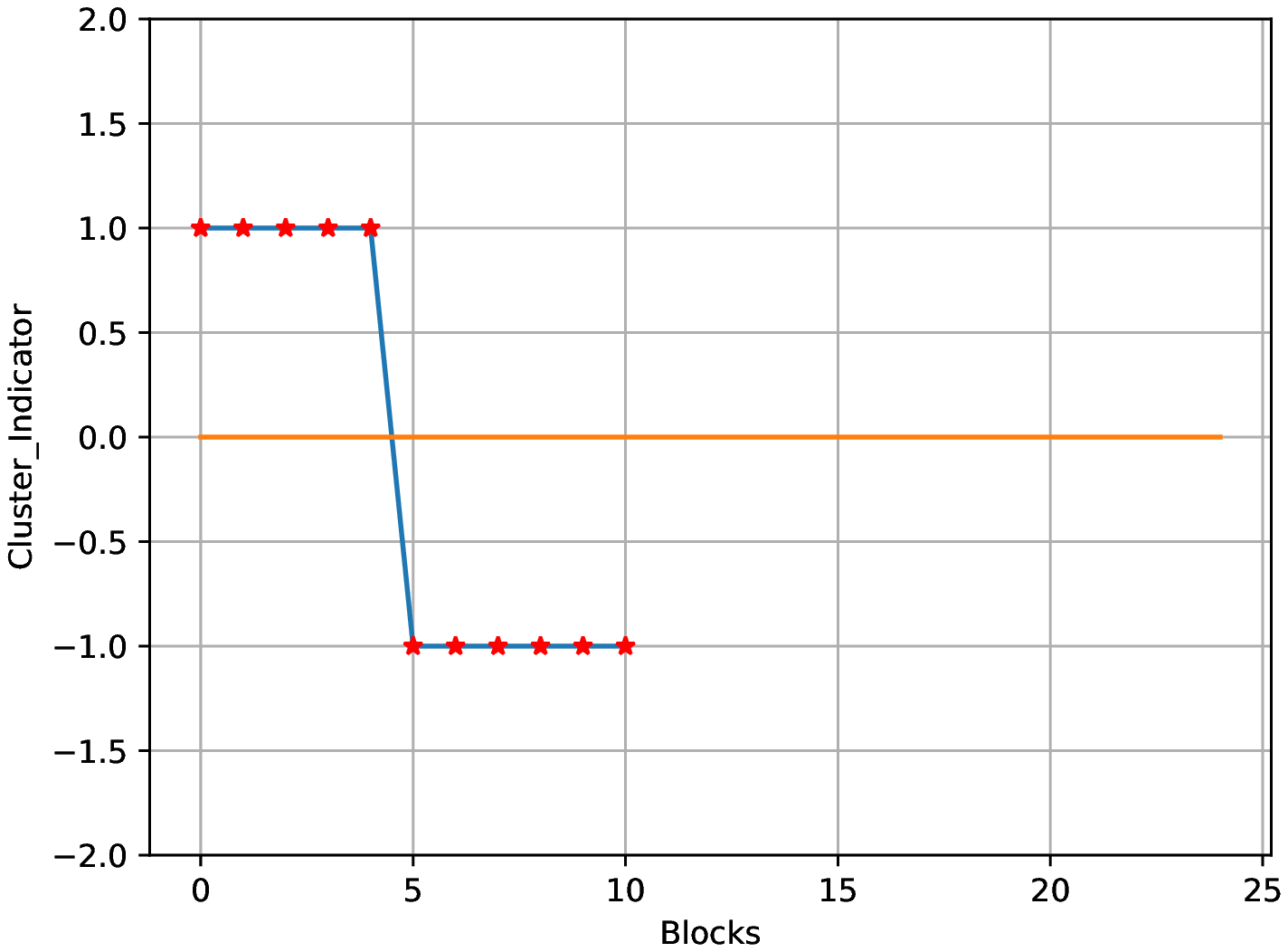}
    \caption{At height = 3}
    \label{fig:h_3}
  \end{subfigure}
  \hfill
  \begin{subfigure}[b]{0.45\columnwidth}
    \includegraphics[width=\linewidth]{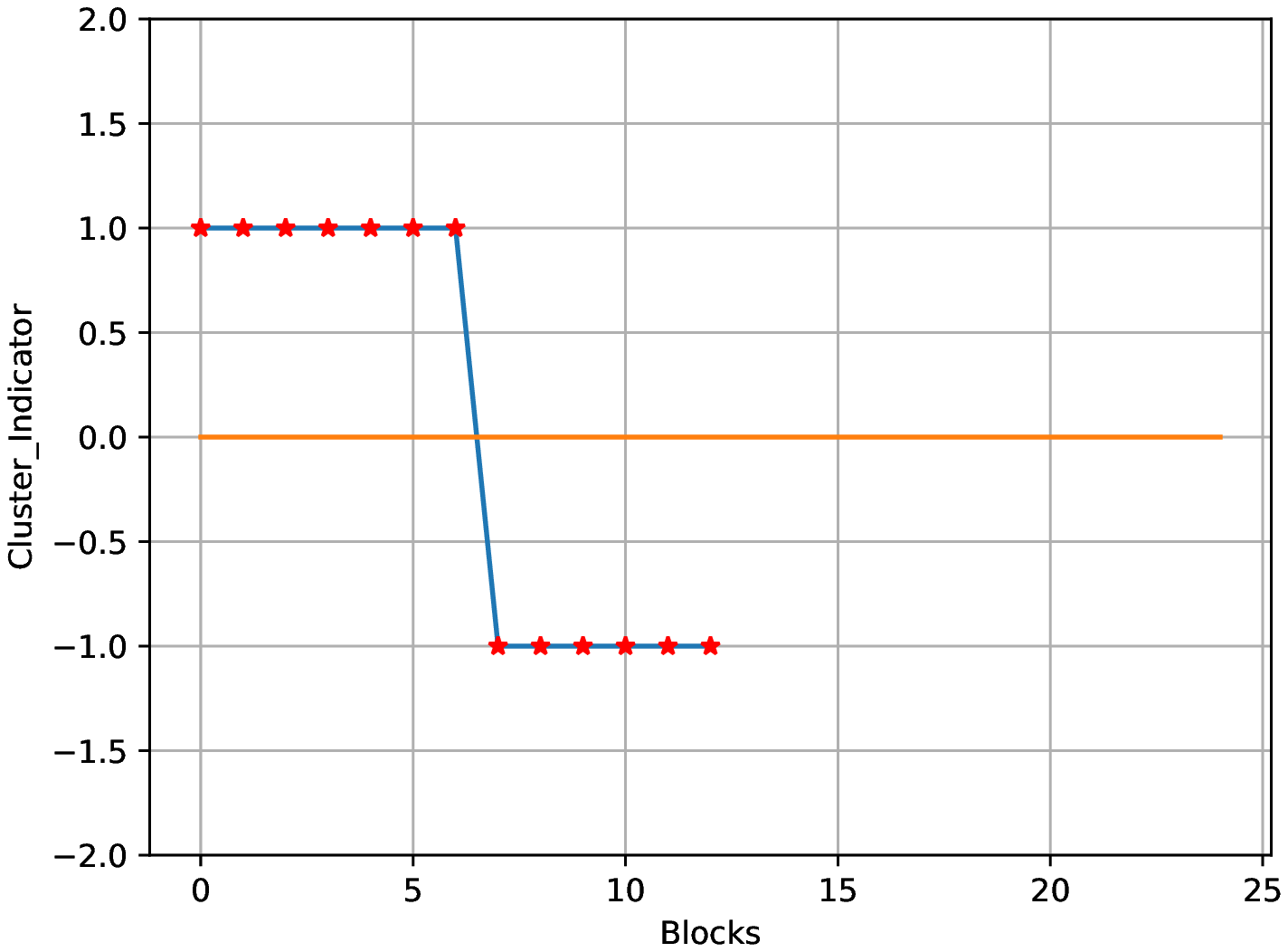}
    \caption{At height = 4}
    \label{fig:h_4}
  \end{subfigure} \\
  \begin{subfigure}[b]{0.45\columnwidth}
    \includegraphics[width=\linewidth]{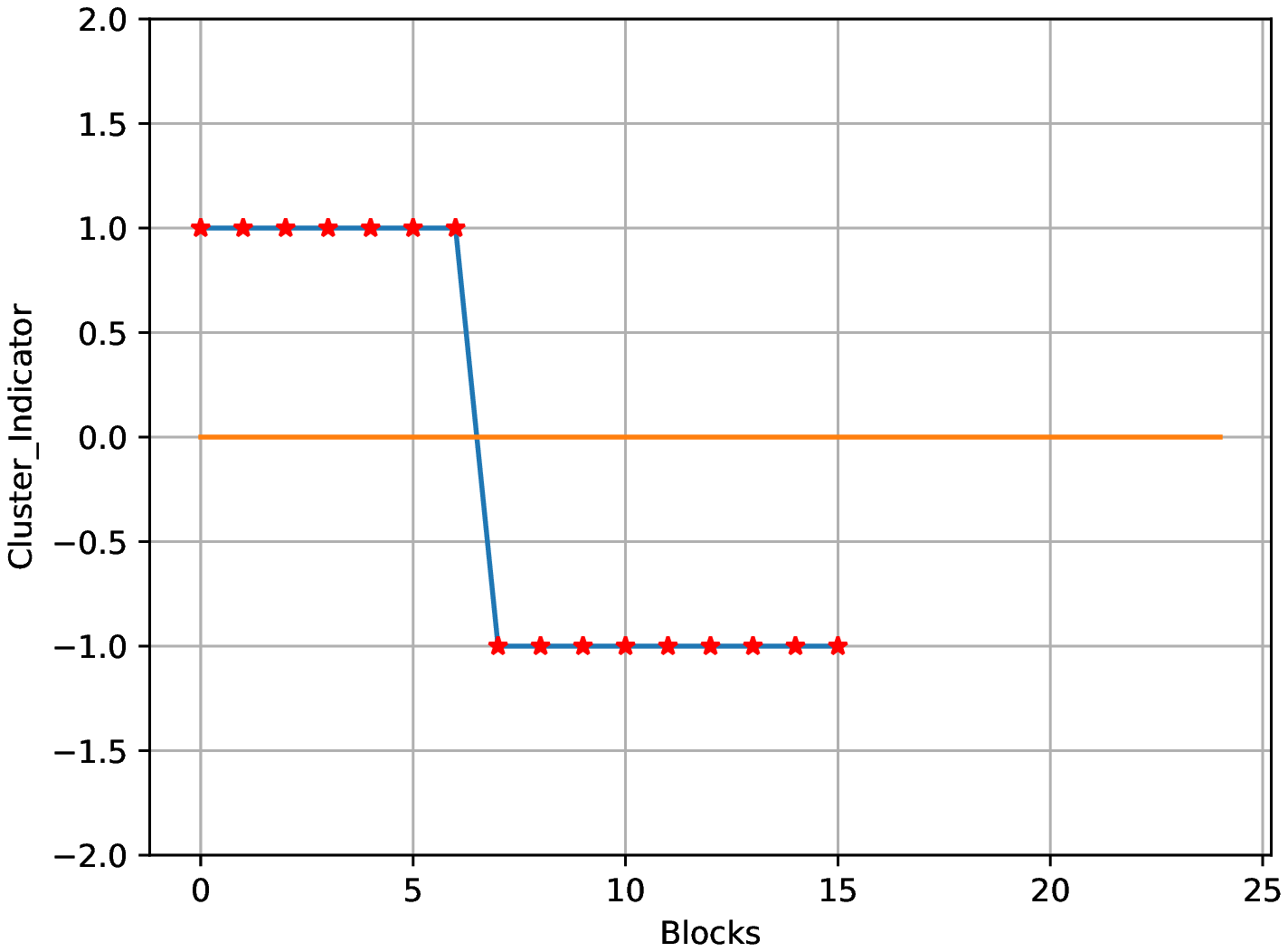}
    \caption{At height = 5}
    \label{fig:h_5}
  \end{subfigure}
  \hfill
  \begin{subfigure}[b]{0.45\columnwidth}
    \includegraphics[width=\linewidth]{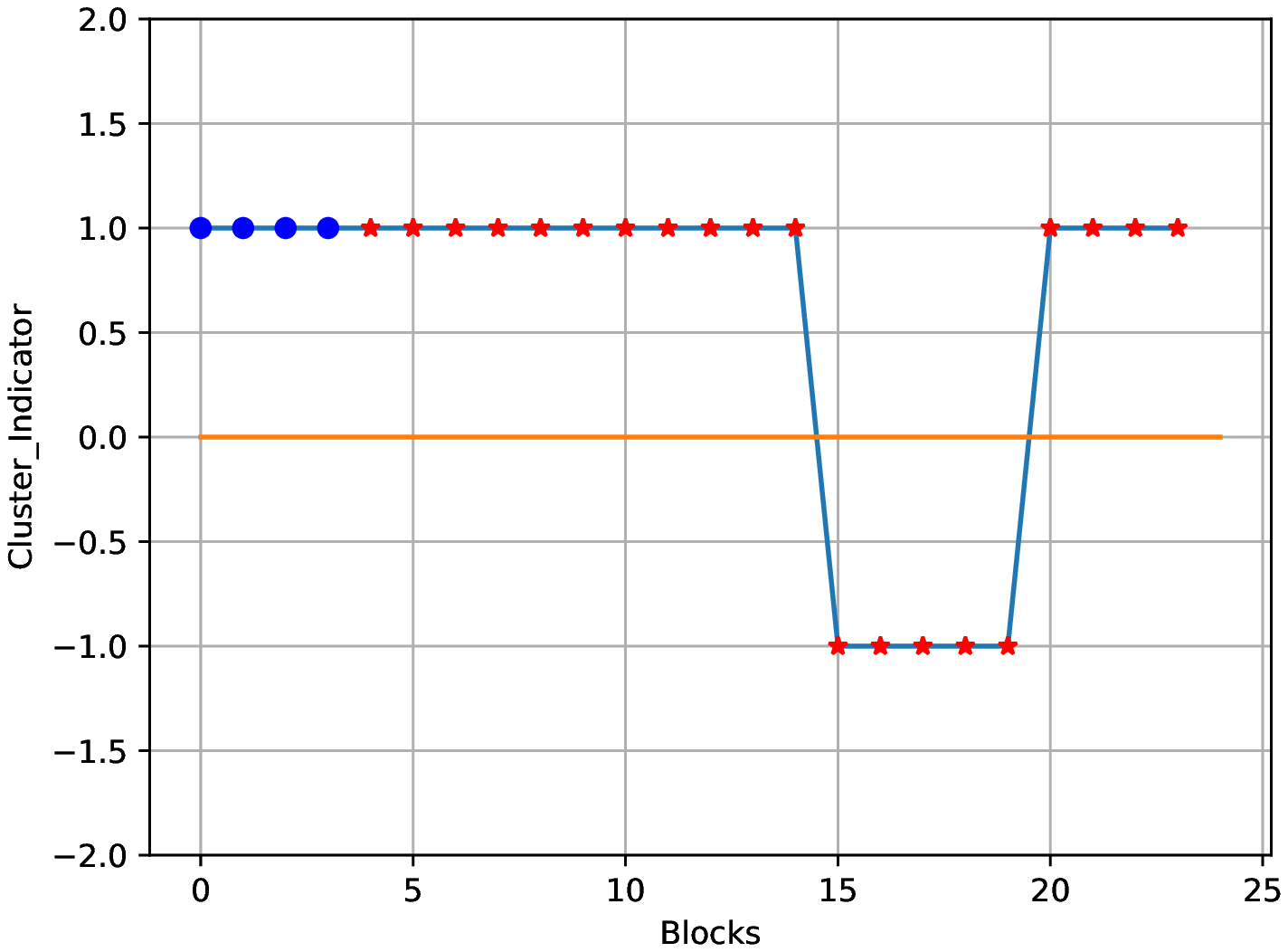}
    \caption{At height = 6}
    \label{fig:h_6}
  \end{subfigure} \\
  \begin{subfigure}[b]{0.45\columnwidth}
    \includegraphics[width=\linewidth]{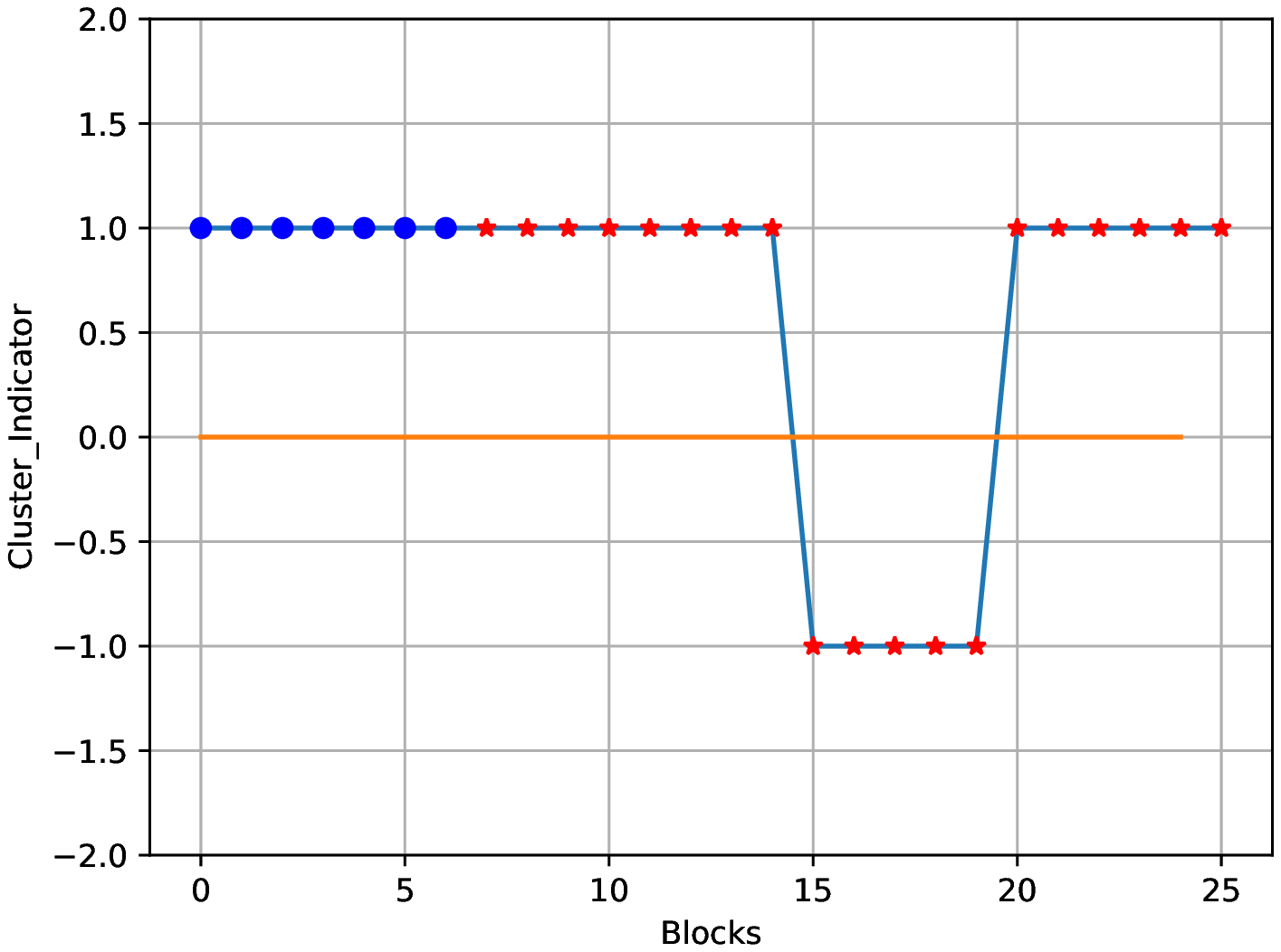}
    \caption{At height = 7}
    \label{fig:h_7}
  \end{subfigure}
  \caption{Spectral properties for $2-clusters$ of blockDAGs shown in Fig. \ref{fig:algo_2}.}
  \label{fig:algo_2_spectral}
\end{figure}

Algorithm 2 is illustrated with the following example shown in Fig. \ref{fig:algo_2} and Fig.\ref{fig:algo_2_spectral}.
In this example, we assume the number of confirmations $k=4$ is constant and leave the analysis of $k$ to the next subsection. The attacker creates a conflicting transaction to a transaction in block $\mathbf{3}$ and includes it in block $\mathbf{15}$. The attacker keeps the blocks $\mathbf{15-18}$ secret until the blocks at $height = 1 $ (blocks $\mathbf{1-3}$) attain
k = 4 number of confirmations by the other clients in the network and broadcast all the blocks from
$\mathbf{15-19}$  whenever the other clients reached $height=6$. Each client will confirm the blocks at a particular height $N$ to either of the clusters $C_1$ or $C_2$ whenever it noticed the height increases to $N+k+1$ in $G_t^c$. For example, a client confirms the blocks at $height=1$, when it reaches  $height=6$. Here, at $height=6$, the blocks $\mathbf{1-3}$ are confirmed to $C_1$ (shown in blue), and block $\mathbf{15}$ is added to $C_2$ (shown in red). At $height=7$, the client excludes blocks $\mathbf{1-3}$ and $\mathbf{15}$ from $G_t^c$ (As per line $2$ of Algorithm \ref{Algorithm2}). Here, the blocks $\mathbf{4-6}$ are added to $C_1$, and block $\mathbf{16}$ is added to $C_2$.

Fig.\ref{fig:algo_2_spectral} shows the spectral properties (The eigenvector corresponding to second smallest eigenvalue) of the blockDAG $G_t^c$ structures shown in Fig. \ref{fig:algo_2}, which illustrates assigning the blocks to clusters $C_1$ (above the orange line) and $C_2$ (below the orange line). Fig. \ref{fig:h_6} and \ref{fig:h_7} show the  blocks $\mathbf{1-6}$ shown in blue colour are added to $\mathbf{blueList}$ and blocks ($\mathbf{15-16}$) are separated from the blocks created by the honest nodes.
\subsubsection{Ordering of confirmed blocks}
Finally, the blocks added to the $\textbf{blueList}$ can be arranged in topological order ($\textbf{ordList}$) as described in Algorithm \ref{Algorithm3} begins with initializing an empty topological queue (\textit{topo\_ q}) and empty ordered list ($\textbf{ordList})$.

\begin{algorithm}[!b]

  \caption{Ordering of blocks}\label{Algorithm3}
  \textbf{Input:} $G_t^c$, $\textbf{blueList}$  \\
	\textbf{output:} $\textbf{ordList}$ - an ordered list of blocks
  \begin{algorithmic}[1]
    \Procedure{ord}{$G^c_t$,$\textbf{blueList}$}
      \State Intialize empty queue \textit{topo\_ q} and empty $\textbf{ordList}$
      \State \textit{topo\_ q.push}(\textit{genesis})
      %\State $\textbf{blueList} \gets {FIND-LIST}(G_t^c, k, N)$
      \While{\textit{$topo\_ q \neq \phi $}} 
          \State $B \gets topo\_q.pop() $
          \State $\textbf{ordList}.add(B)$
          \State $children(B) \gets \{j: (j,B) \in E\} $
          
          \State Sort($children(B)$) based on their $hash$ values
          \For{all $C \in children(B) \cap \textbf{blueList}$}
             \State $topo\_q.push(C)$
		  \EndFor
      \EndWhile
      \State \textbf{return} $\textbf{ordList}$ 
    \EndProcedure
  \end{algorithmic}
\end{algorithm}

\begin{itemize}
\item Each client add the blocks in $\textbf{blueList}$ to $topo\_q$ starting from the $genesis$ block.

\item Arrange the $children$\footnote{children of a block $\mathbf{B}$ are the blocks that refer to $\mathbf{B}$ directly in the blockDAG.} of a block in ascending order of their hash value, which indicates the block with a lesser hash value (block with smaller hash  than the target value in PoW puzzle \cite{bitcoin}) added first among children of $\mathbf{B}$ with the same block height.
\end{itemize}

When a block $\mathbf{B}$ is pushed into \textit{topo\_ q}, all blocks referenced by $\mathbf{B}$ are also pushed. The algorithm adds all the blocks and their children in $\mathbf{blueList}$ to \textit{topo\_ q}. The client $c$ executes Algorithm \ref{Algorithm3} in $\mathcal{O}(|\mathbf{blueList}|+|E|)$.
\subsubsection{Transaction security}
Consider a transaction $tx \in \mathbf{B}$, where $\mathbf{B}$ is in $\mathbf{blueList}$ of $G_t^c$. In order to make $tx$ invalid, an attacker creates a conflicting transaction $tx'$ in a block $\mathbf{C}$, which is a part of the attacker's secret chain similar to the secret chain shown in Fig. \ref{fig:attacker_model}. The recipient of the $tx$ wait for $k$ number of confirmations to find the $\mathbf{blueList}$ around the $tx$ and eliminate the block having $tx'$ as invalid. Thus, the implication of block security through $k$ is also implicated to transaction security.

\subsection{Mathematical Analysis of UL-blockDAG}
The block creation rate $\lambda$ is very large for creating the blockDAG structure of ledger. So, the UL-blockDAG protocol does not rely on the condition  $\lambda D << 1$ which is imposed on longest chain rule (bitcoin) and optimal longest chain rule protocols.
\subsubsection{Transaction throughput (TPS)}
\begin{lemma}
\label{lemma:beta_dag}
For a UL-blockDAG protocol based blockchain network generating blocks at a rate $\lambda$ and delay diameter $D$, the lower bound on the main chain growth rate $\beta$ for a small $\delta$ ($\delta << 1$) is
\begin{equation}
%\beta \geq \frac{2\lambda D \sqrt{N}}{\sqrt{N} - 3}
\beta  \geq \frac{\lambda D}{1 + D - \frac{3+\delta}{\sqrt{N}}}
\label{eq:beta_dag}
\end{equation}
\end{lemma}
\proof See Appendix C.
\begin{corollary}
For $N\to\infty$
\begin{equation}
\label{eq:beta_dag_inf}
\beta \geq \frac{\lambda D}{1 + D}.
\end{equation}
\end{corollary}
\begin{lemma}
The minimum TPS in blockDAG ledger based blockchain network is
\begin{equation}
TPS(\lambda,b) = \frac{\lambda D}{1+D}  b K \approx \lambda b K
\end{equation}
\end{lemma}
\subsubsection{Number of confirmations (k)}
\textbf{Definition 6. {\textit{future}(B)}.} The \textit{future} of a block $\textbf{B}$ is a set of all the blocks which refer to block $\textbf{B}$ directly or indirectly, which were created after the block $\textbf{B}$. The spectral clustering algorithm is based on intra-cluster and inter-cluster references depend on the \textit{future} of a block till the block attain the required number of confirmations $k$. Similarly, $future^i(\textbf{B})$ is denoted as a set of blocks that refer to $\textbf{B}$ directly or indirectly at height $i$ in $G_t^c$ and $future^i(\textbf{B}) \subseteq future(\textbf{B})$. \\
\textbf{Definition 7. {Risk(\textbf{B},k)}.} Let $\textbf{B}$ and $\textbf{B}'$ are honest and attacker blocks respectively having conflicting transactions, the risk of including $\textbf{B}'$ instead of $\textbf{B}$ in $\textbf{ordList}$ is defined as
\begin{equation}
Risk(\textbf{B},k) = Pr\left[\sum_{i=N+1}^{N+k+1}R_{h,i} < R_{a,N+k+1}\right]
\label{eq:risk}
\end{equation} 
Where,
\begin{itemize}
\item \begin{equation}
R_{h,i} = \big|future^i(\textbf{B})\big|
\end{equation} 
\item  \begin{equation}
\sum_{i=N+1}^{N+k+1}R_{h,i} = \Bigg|  \left \lbrace \bigcup_{i=N+1}^{N+k+1} future^i(\textbf{B}) \right\rbrace \Bigg|
\label{eq:R_h}
\end{equation} 
\item \begin{equation}
R_{a,N+k+1} =  \Big| future^{N+k+1} (\textbf{B}')\Big|
\label{eq:R_a}
\end{equation}
\end{itemize}
\begin{proposition}
\label{prop:Risk}
Suppose, a block $\mathbf{B}$ included in the DAG $G_t^{c}$ at time $t$ ($\textbf{B} \in G_t^{c}$) and at any time $s >> t$ ($\textbf{B} \in G_s^{c}$)  
\begin{equation}
\lim_{k \to \infty} Risk(\textbf{B},k) = 0
\end{equation}
\end{proposition}

\begin{lemma}
For any $\epsilon > 0$ ($\epsilon << 1$), $Risk(\textbf{B},k) < \epsilon$, the minimum number of confirmations required for the block $\textbf{B}$ is 
\begin{equation}
\frac{3(\lambda D + 1)(1-\epsilon)}{4\left[\lambda (1-q)D + 1\right]}
\label{eq:k}
\end{equation}
\label{lemma:confirmations}
\end{lemma}
\proof See Appendix D.

\subsubsection{Safety and Liveness Properties of the proposed blockDAG consensus protocol}
Proposition \ref{prop:Risk} essentially gives the Safety and Liveness properties required for the distributed consensus protocol. Indeed, once $Risk(\mathbf{B},k) < \epsilon$ (based on the minimum number of the confirmations $k$ required for the block $\mathbf{B}$) is satisfied, the client can decide to accept transactions in $\mathbf{B}$ (Liveness). The accepted transactions are guaranteed to be irreversible (Safety) with a probability of $\epsilon$ as confirmations to $\mathbf{B}$ increase with time.

\subsection{Suitability for Smart Contract applications}
Smart contracts \cite{smartContracts} are the self-executable computer codes deployed on the blockchain with a set of functions and state variables. These smart contracts found applications in Health care \cite{medical}, IoT services \cite{IoT}, and so forth.  

The execution of smart contract functions changes the values of one or more state variables. The change of state variable values create transactions on the blockchain. These transactions included in the blocks construct a DAG structure called blockDAG. The transactions in a block are ordered as per their order of arrival. To maintain the consistency of the state variable values,  the transactions related to that smart contract should be arranged in linear order. The topological ordering algorithm
(Algorithm 3) has been used to organize the blocks in a linear order to make the protocol suitable for smart contract applications. The IoT services require a very low operational latency in processing the transactions of the blockchain. The minimum number of confirmations as per \eqref{eq:k} shows the confirmation delay is less in the proposed protocol.

\subsubsection{The effect of transaction arrival rate on the security}
In some IoT systems, the transaction arrival rate is fluctuant. In the proposed UL-blockDAG protocol, the two independent parallel processes create the blocks with  rates $\lambda$ and $\lambda D$ (as per the Lemma \ref{lemma:beta_dag}). In such a scenorio, the blocks are created with lesser block size for lower transaction volume and the blocks are created with larger block size  for higher transaction volume. Thus, the transaction arrival rate does not affect the block arrival rate. 

The confirmation delay $k$ which characterizes the double-spend attack depends on the number of block references to $\mathbf{B}$ in $future(\mathbf{B})$ and also the number of references to a block $\mathbf{B}'$ (having conflicting transaction with a transaction in $\mathbf{B}$) in $future^{N+k+1}(\mathbf{B}')$. Moreover, $future(\mathbf{B})$ and $future^{N+k+1}(\mathbf{B}')$ depends on the two independent processes with rates $\lambda$ and $\lambda D$ (as per Lemma \ref{lemma:beta_dag}), but not on the transaction arrival rate. So, the fluctuation in the transaction arrival rate neither affects the $k$ nor the security of the system.
\section{Results and Discussion}
\subsection{Event-driven simulation}
The event-driven simulator generates the events as per the information propagation protocol \cite{info} for bitcoin protocol with inputs - number of nodes ($n$), block creation rate ($\lambda$), Genesis block ($genesis$), Hash rate distribution ($H$) \cite{hashrate} and duration of the simulation ($Sim\_Time$). The simulator generates the events for creating a block, broadcasting the block to neighbors, and adding the block after verifying the block height and hash of the previous block. 

\begin{algorithm}[!t]
  \caption{Event-driven Simulator}\label{Algorithm4}
  \textbf{Input:} $n$ - number of nodes, $\lambda$ - Block creation rate, $Sim\_Time$ - Duration of the simulation, $genesis$ - Genesis Block, $p_c$ - Fraction of hash rates of each node   \\
	\textbf{output:} $blockLedger$ 
  \begin{algorithmic}[1]
    \Procedure{runSim}{$n$,$\lambda$,$Sim\_Time$,$genesis$}
      \State Initialize empty $event\_q$ and $miners$
      \State $t = 0.0$ 
      \State Initialize the $miners$ with inputs and $event\_q$   \Comment It also \\ \hspace{2.5cm} add a new $mine\_block$ event to $event\_q$
      \State $add\_peers(8)$     
      \While{\textit{$ Sim\_Time < t$}} 
          \State $t,t\_event \gets event\_q.pop() $
          \State $receive\_event(t,t\_event)$
       \EndWhile
      \State \textbf{return} $blockLedger$ 
    \EndProcedure
  \end{algorithmic}
\end{algorithm}

\begin{algorithm}[!b]
  \caption{Generation of Events}
  \label{Algorithm5}
  \textbf{Input:} $t$ - Time of occurrence of an event, $t\_event$ - Event to be executed at time $t$   \\
	\textbf{output:} Update  blockLedger 
  \begin{algorithmic}[1]
    \Procedure{genEvents}{$t$, $t\_event$}
    	\If{$t\_event.action ==$ "block"}
			\State Verify($t\_event.payload$)
			\State Generate $add\_block$ event
		\ElsIf{$t\_event.action == $"addblock"}
			\State $blocks.append(t\_event.payload)$
			\State $chain\_head.append(hash(t\_event.payload))$	
			\State Announce($t\_event.payload$)
			\State generate $inv$ events 
			\State $t \gets t + random.exp\left(\frac{1}
				{\lambda \times p_c}\right)$  
			\State Generate next $mine\_block$ event at $t$
		\ElsIf{$t\_event.action == $"inv"}
			\If{$t\_event.payload$ not in $blocks$} 
				\State Request($t\_event.payload$)
				\State Generate $get\_block$ event
			\EndIf 
		\ElsIf{$t\_event.action ==$ "$get\_block$"} 
				\State Send($t\_event.payload$)
		    \EndIf

          \EndProcedure
  \end{algorithmic}
\end{algorithm}
We randomly generate eight peers for each client $c$, similar to the bitcoin P2P network. The timing information for mining a block event of a miner $c$ is drawn from an exponential distribution with mean $\lambda \times p_c$. All the event information stored in a time  priority queue named as $event\_queue$. The ledger data stored in a dictionary named $blocks$ with $blockhash$ as a key and $block$ data structure as a value, and the $chain\_head$ stores the list of block hashes. These two data structures were used in the analysis of the protocol. Algorithm \ref{Algorithm4} describes the execution of events from the $event\_queue$ on a time priority basis.

The events are divided into four types as per the information propagation protocol described in \cite{info} of bitcoin for propagating a block from miner to reach the entire network. The events are classified as \textit{inv} -  sending a new block hash invitation, \textit{getblock} - requesting a new block, \textit{block} - sending a block to its peers and \textit{addblock} -  adding a received block to its local copy of the blockchain. After adding a mined/received block, each miner starts mining with exponentially distributed time intervals with a rate of their hash rate (line $9$ of Algorithm \ref{Algorithm5}). The generation of these events is described through Algorithm \ref{Algorithm5}. 
The payload of an event is defined as
\begin{align}
t\_event.payload &= \begin{cases}
    hash(block), & \text{for inv event}.\\
    block , & \text{for other events}.
  \end{cases}
  \label{eq:t_event}
\end{align}

\begin{algorithm}[!t]
  \caption{Modified $addblock$ event for UL-blockDAG}\label{Algorithm6}
  \textbf{Input:} $t$ - Time of occurrence of an event, $t\_event$ - Event to be executed   \\
	\textbf{output:} $\textbf{ordList}$ 
  \begin{algorithmic}[1]
    \Procedure{$addblock$}{$t$, $t\_event$}
    		\State $blocks.append(t\_event.payload)$
			\State $chain\_head[bloc\_height].append(hash(t\_event.payload))$
			\State Mapping(hash($t\_event.payload$)) 
			       \Comment Mapping \\ \hspace*{0.5cm}   
			        block hashes to natual numers and 
			construct DAG  suitable \hspace*{0.5cm} for 
			consensus algorithm execution
			\State $\textbf{ordList} \gets$ UL-blockDAG consensus Algorithm
			\State Announce($t\_event.payload$) 	
			\State generate $inv$ events 
			\State $t \gets t + random.exp\left(\frac{1}
				{\lambda\times p_c}\right)$ 
			\State Generate next $mine\_block$ event at $t$
			\State \textbf{return} $\textbf{ordList}$
      \EndProcedure
  \end{algorithmic}
\end{algorithm}
The simulation Algorithms are similar for blockDAG protocol, except in constructing ledger ($addblock$ event) and consensus algorithm. The $chain\_head$ for blockDAG structure is a dictionary  with block height as key and block hashes at that height as values. Algorithm \ref{Algorithm6} describes the modified $addblock$ event for blockDAG protocol simulation.

%%%%%
\subsection{Comparison of mainchain block growth rate with simulation results}
TABLE \ref{table:values} lists the values of the parameters used for generating the results in this section.  See   TABLE \ref{table:symbols} for a description of the symbols.

\begin{table}[!b]
\caption{\\ Parameter values for the case of Bitcoin}
  \label{table:values}
\centering
\begin{tabular}{c l}
\hline
    \textbf{Parameter}&\textbf{value}\\
\hline
    $n$ & $100$   \\
    $N_t$  & $8$ \quad \cite{bitnodes}  \\
    $P_e$  & $8.0/(n-1) \approx 0.08$ \\
	$T_p$& $30$ msec \\
	$b$ & $4$ MB \\
	$R$ & $>$ $25$ Mbps \quad \cite{upload} \quad but chosen $10$ Mbps\\
$K$ & $4$ txn's/KB \quad for bitcoin \cite{blockchain}  \\ 
		\hline
		  \end{tabular}
\end{table}

We have conducted an event-driven simulation using python by generating events for one day as per the information propagation protocol in \cite{info} for bitcoin blockchain network with $n = 100$ nodes, and $13$ miners having the Hashrate distribution shown in \cite{hashrate}. The simulation are conducted such that delay diameter should be equal to $D \approx 10 sec $ calculated for parameter values shown in TABLE \ref{table:values} using \eqref{eq:D}. 

\begin{figure}[!t]
    \centering
    \begin{subfigure}[t]{0.45\textwidth}
        \centering
        \includegraphics[width=\columnwidth]{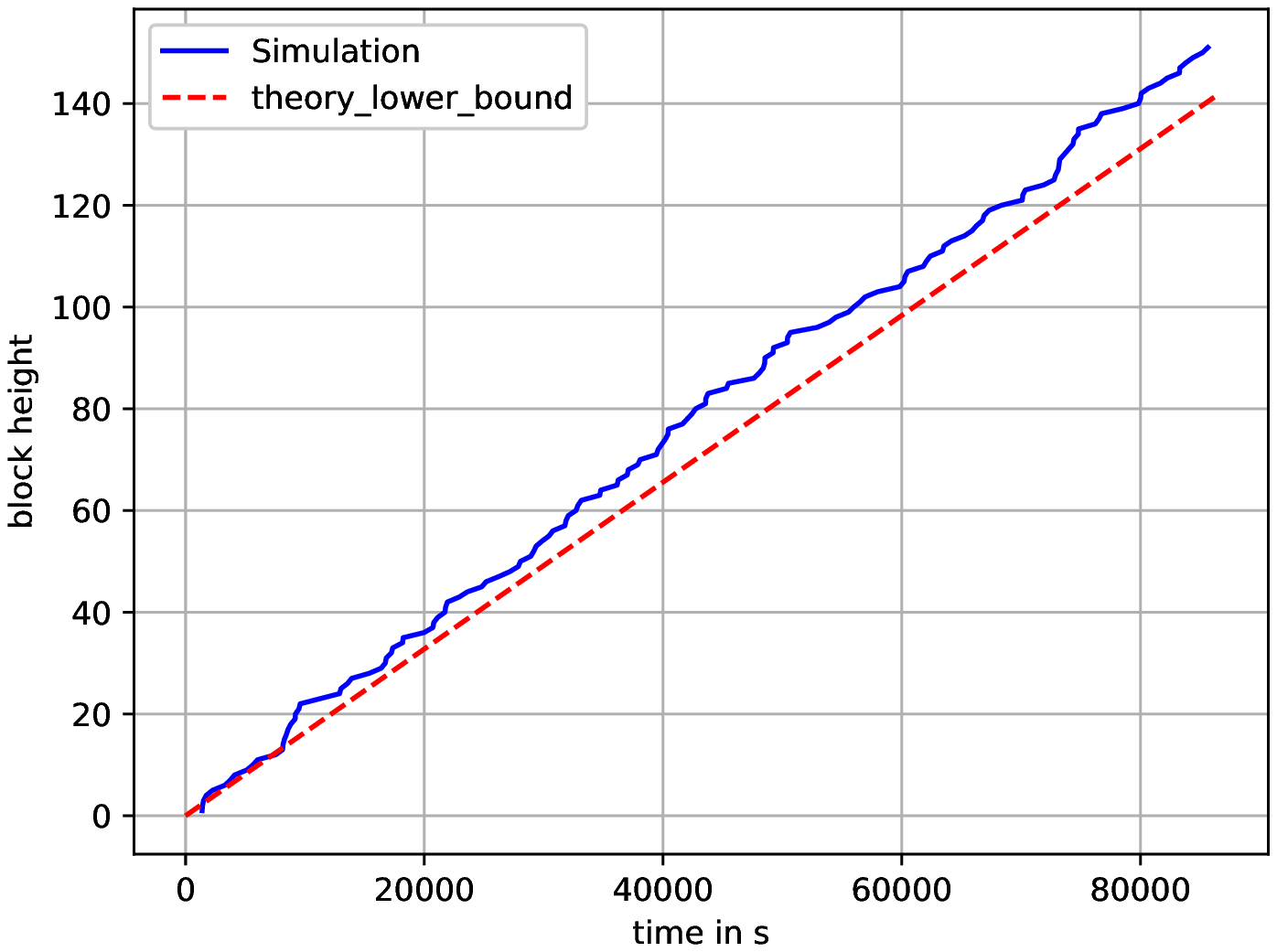} 
        \caption{$\lambda=\frac{1}{600}$ (Bitcoin)} \label{fig:block_height_600}
    \end{subfigure}
    \hfill
    \begin{subfigure}[t]{0.45\textwidth}
        \centering
        \includegraphics[width=\columnwidth]{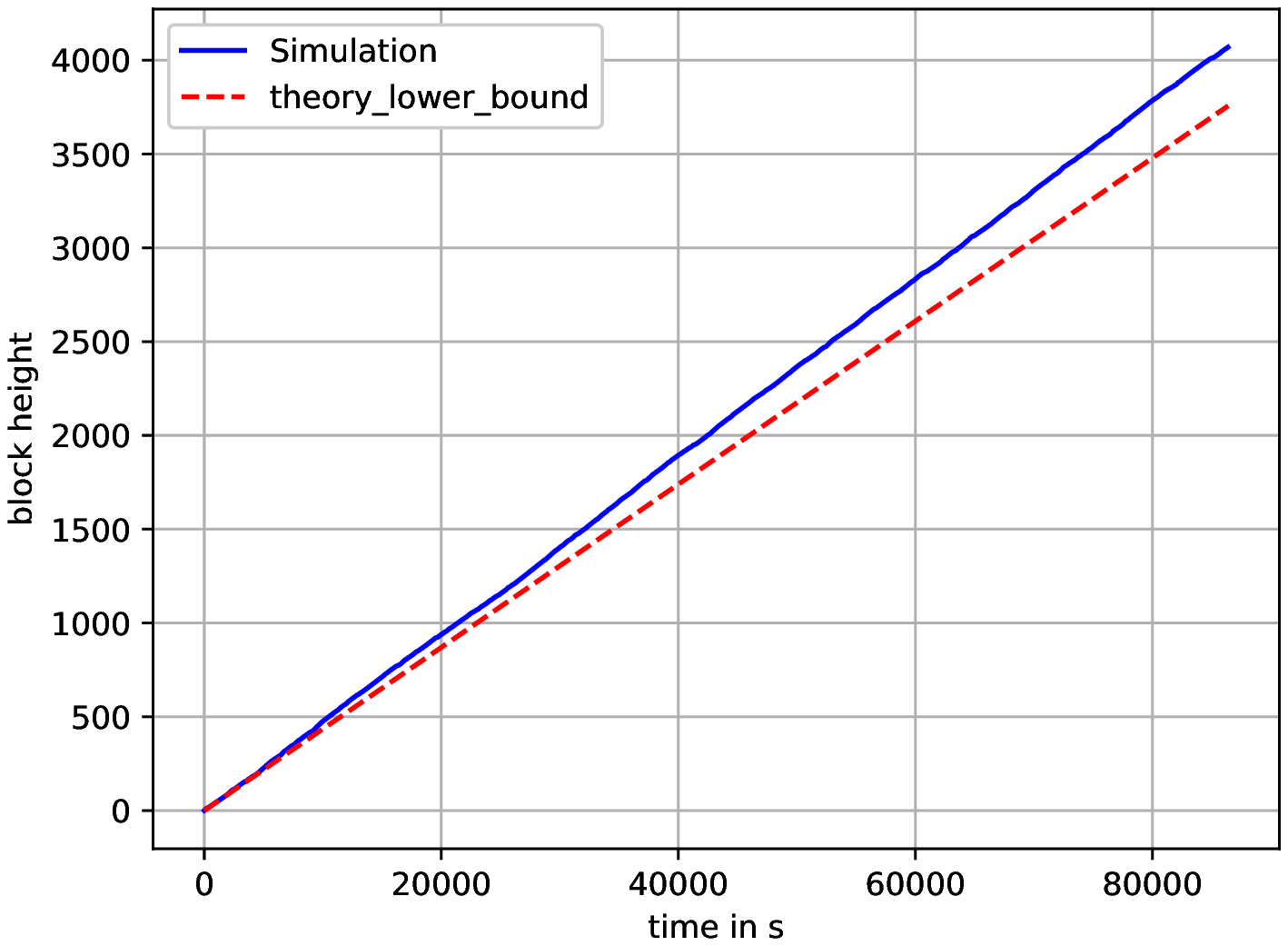} 
        \caption{$\lambda=\frac{1}{12}$ (Optimal longest chain rule)} \label{fig:block_height_optimal}
    \end{subfigure}
    \hfill
	\begin{subfigure}[t]{0.45\textwidth}
        \centering
        \includegraphics[width=\columnwidth]{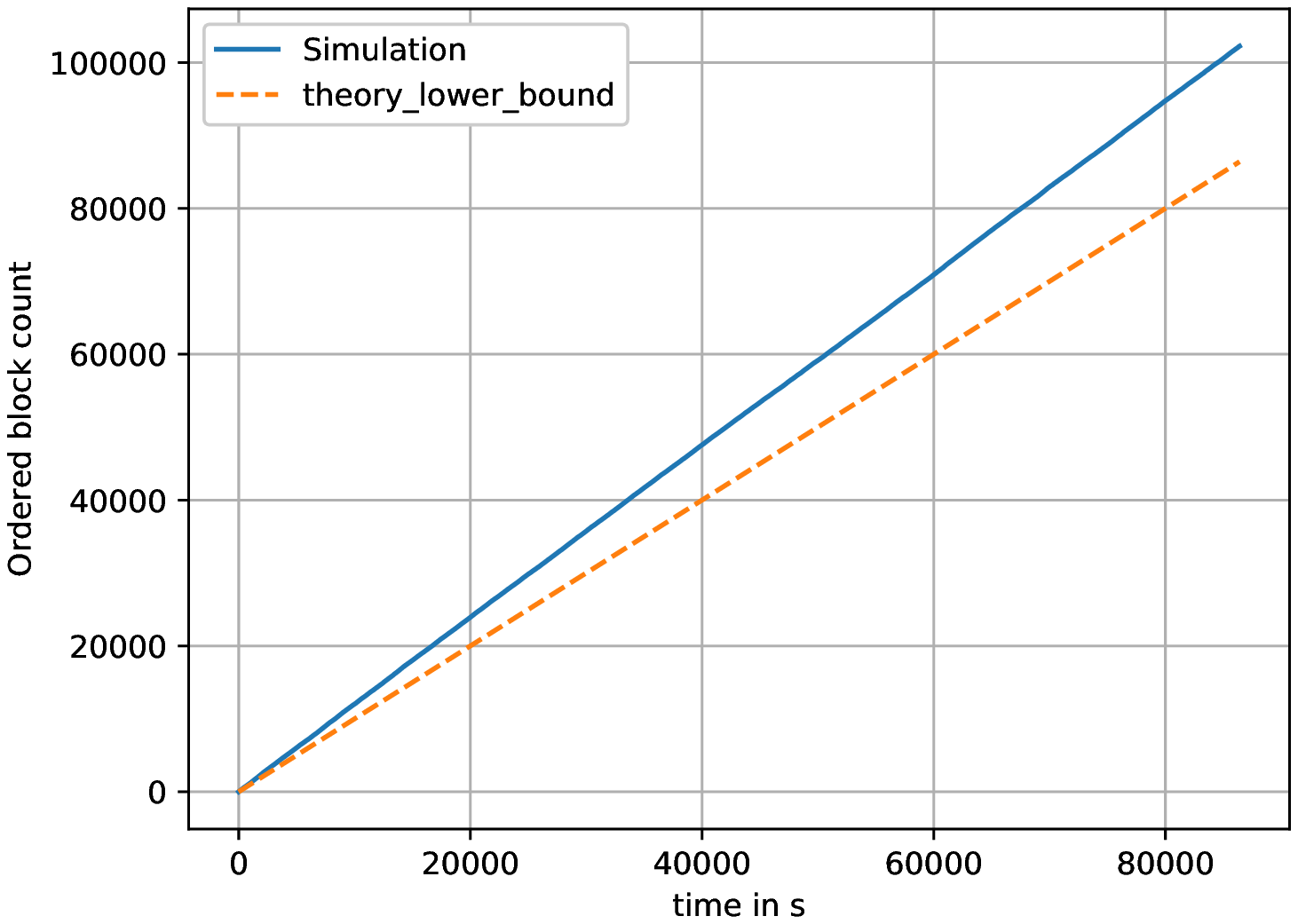} 11
        \caption{$\lambda=1$ (UL-blockDAG protocol)} \label{fig:dag_blocks}
    \end{subfigure}
    \caption{Time in sec vs Number of blocks  for a duration of $1$ day.}
\end{figure}

We have chosen the optimal block creation rate $\lambda = \frac{1}{12}$ ($ \frac{1}{\lambda} > D $). Fig. \ref{fig:block_height_600} and Fig. \ref{fig:block_height_optimal} show the block height w.r.to time of creation. In both cases, the main chain growth rate of simulations is satisfied the theoretical lower bound \eqref{eq:beta_inf} for longest chain rule consensus protocol. Fig. \ref{fig:block_height_optimal} also shows the significant increase in TPS to $\approx 4000$ per day.

The simulations carried out for UL-blockDAG protocol  with $\lambda = 1$ block/sec (or block interval is $1$ sec).  Fig. \ref{fig:dag_blocks} shows block growth rate w.r.to to time and compared it with the theoretical lower bound \eqref{eq:beta_dag_inf}. The results show an increase in block growth rate with approximately a block per every second due to blockDAG structure of ledger and the security guaranteed by the UL-blockDAG protocol.

%\twocolumn
\subsection{Correctness of Algorithm \ref{Algorithm2} through simulations}

\begin{figure}[!t]
  \begin{subfigure}[t]{0.45\textwidth}
    \includegraphics[width=\linewidth]{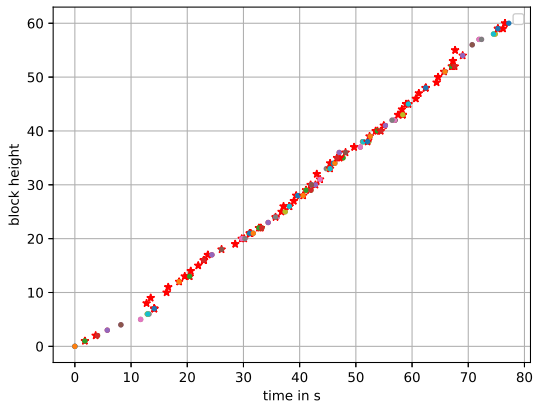}
    \caption{$q = 0.33$}
    \label{fig:30}
  \end{subfigure}
  \hfill %%
  \begin{subfigure}[t]{0.45\textwidth}
    \includegraphics[width=\linewidth]{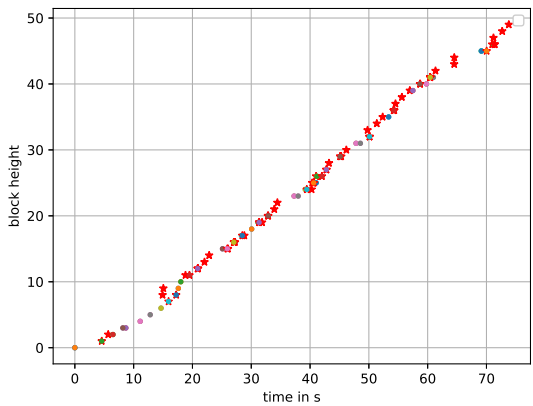}
    \caption{q = 0.51}
    \label{fig:50}
  \end{subfigure}
\caption{All the blocks created by the miners in a duration of 86 seconds. Attacker blocks are shown
in red color, and other miners' blocks are shown
in different colors.. Blocks created by the attacker from height $\mathbf{3-7}$ are not included in $\textbf{blueList}$. The coincidence of the blocks
created by different miners (with different colored blocks) at each block height indicates multiple blocks at each height due to the ledger's DAG structure.}
\label{fig:attacker_honest}
\end{figure}

We generated the events such that the attacker with $33.33\%$ (and $51 \%$) of hash rate follows the double-spending strategy similar to an example shown in Fig. \ref{fig:algo_2} from $height = 3$ to $height = 7$ by choosing  $k=5$. Fig. \ref{fig:attacker_honest} shows a part (for a duration of $86$ sec) of total confirmed blocks ($\textbf{blueList}$) created by all miners in the network, where the attacker blocks (red color) from height $\mathbf{3-7}$ are not included in the $\textbf{blueList}$, and only honest blocks are present from height $3$ to $7$ as per Algorithm \ref{Algorithm2}. 
\subsection{The fairness of the proposed models}  
\begin{figure}[!t]
    \centering
    \begin{subfigure}[t]{0.45\textwidth}
        \centering
        \includegraphics[width=\columnwidth]{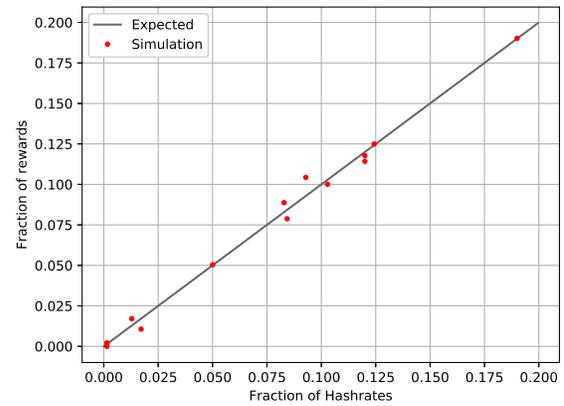} 
        \caption{$\lambda=\frac{1}{600}$ (Bitcoin)} \label{fig:sim_600}
    \end{subfigure}
    \hfill
    \begin{subfigure}[t]{0.45\textwidth}
        \centering
        \includegraphics[width=\columnwidth]{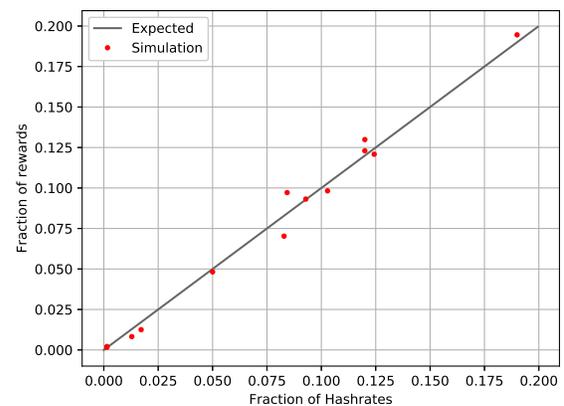} 
        \caption{$\lambda=\frac{1}{12}$ (Optimal longest chain rule)} \label{fig:sim_optimal}
    \end{subfigure}
    
    \begin{subfigure}[t]{0.45\textwidth}
        \centering
        \includegraphics[width=\columnwidth]{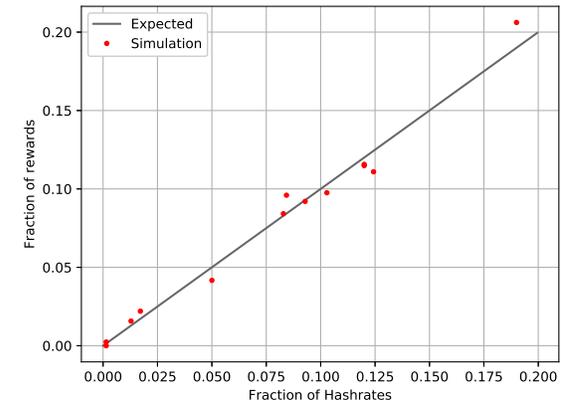} 
        \caption{$\lambda = 1$ (UL-blockDAG)} \label{fig:sim_dag}
    \end{subfigure}

    \caption{Hashrate Vs Rewards proportion for $1$ day.}
\end{figure}

The fairness of both frameworks optimal longest chain rule and UL-blockDAG  are shown in 
Fig. \ref{fig:sim_optimal}, and Fig. \ref{fig:sim_dag}, which shows the proportion of the rewards (shown by red dots) of each miner are nearly equal to their proportion of the hash rates (shown by a line) in the network and is comparable with the rewards proportion for $\lambda = \frac{1}{600}$ shown in Fig. \ref{fig:sim_600}. These results show that the longest-chain rule with an optimal block creation rate of $\lambda = \frac{1}{D}$ and UL-blockDAG protocol with $\lambda = 1$ blockchain networks perform similar to the bitcoin system with block creation rate of $\frac{1}{600}$.

\begin{figure}[!t]
    \centering
        \includegraphics[width=\columnwidth]{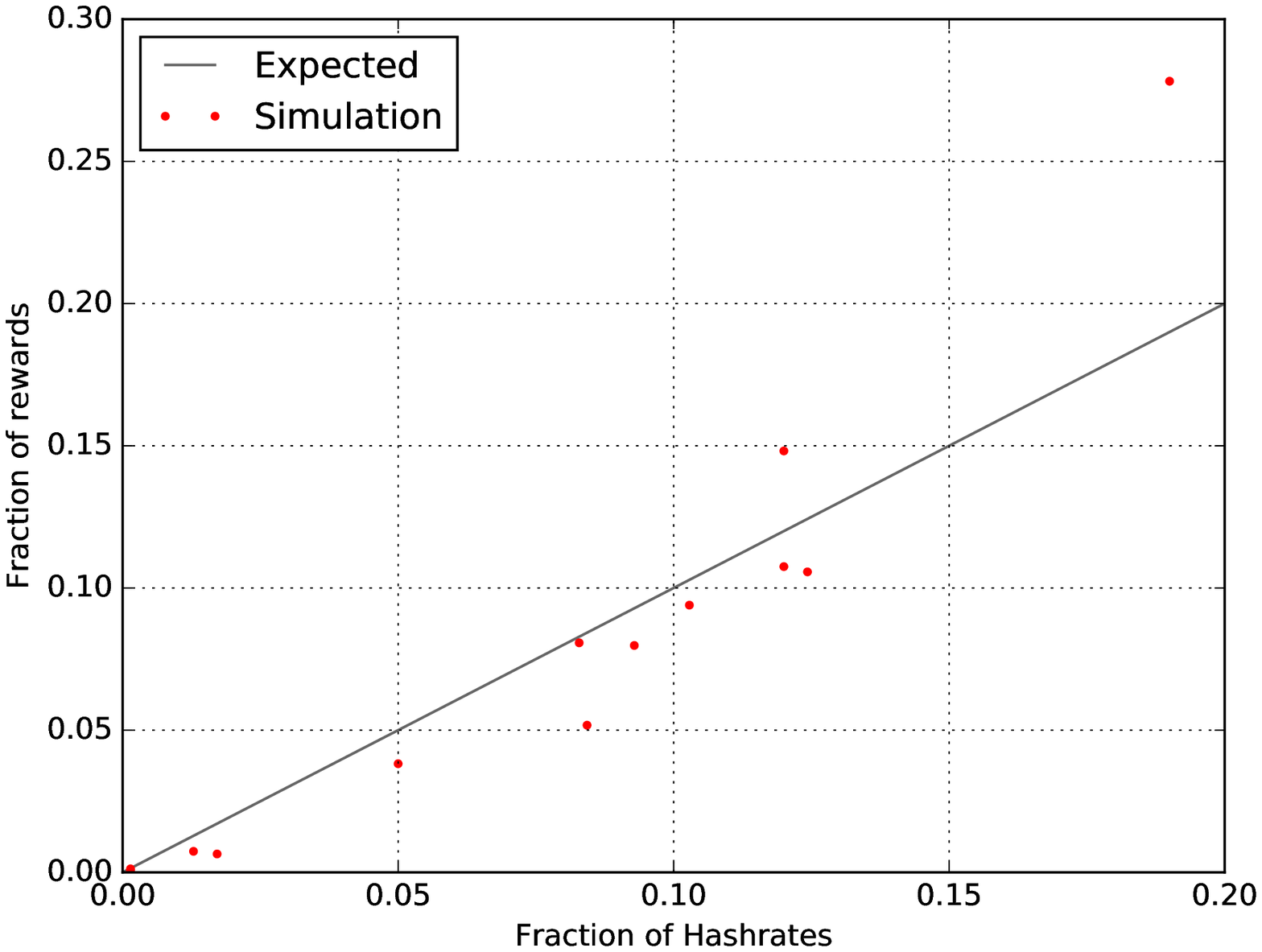} 
        \caption{$\lambda=\frac{1}{5}$ \quad ($\lambda > \frac{1}{D} $)} \label{fig:beyond_optimal}
\end{figure}

Fig. \ref{fig:beyond_optimal} show that in a longest chain rule protocol with the block ceration rate beyond the optimal point \eqref{eq:opt_lam_N},  the miner with the largest hash rate will dominate the system with more rewards than her proportion of the hash rate, resulting in possible double-spend attack by the dominant miner in the network. 

\section{Conclusions and Future Research}
In this paper, we propose two different frameworks for scaling the transaction throughput (TPS) of the PoW based blockchain networks without compromising the fairness of the system. In the first framework, we obtained an analytical expression for the optimal block creation rate and optimal throughput by considering the bitcoin network as Erd$\ddot{o}$s - R$\acute{e}$nyi random network topology. In the second framework, we show the graph clustering based on the spectral graph theory concepts for DAG to separate the blocks created by an attacker with a double-spend strategy. The simulation results show that Algorithms 2 separates the attacker blocks with different attacker hash rates (0.33 and 0.51) when an attacker tries to attempt a double-spend attack. Algorithm \ref{Algorithm3} orders the blocks in a linear order, making the protocol suitable for smart contract applications. In both frameworks, the results show a significant improvement in the transaction throughput and the fairness of the network in terms of the proportion of the miners' rewards and hash rates. The results also compared the simulated block growth rate with the theoretical lower bound derived for both the frameworks.
The theoretical minimum number of confirmations show the lesser confirmation times for the transactions of a block. In the future, we compare the different spectral clustering methods for the proposed consensus protocol for optimizing the computational complexity of the blockDAG consensus protocol.  We also explore the methods to avoid the repeated redundant transactions stored in the blocks at the same height due to the DAG structure.

\ifCLASSOPTIONcompsoc
  \section*{Acknowledgments}
\else
  \section*{Acknowledgment}
\fi
We thank the reviewers for their insightful, helpful, and detailed comments.
This work is supported by 5G Research and Building Next-Generation Solutions for Indian Market Project, Department of Information Technology, Government of India. The
preliminary versions of this paper - "Optimal Transaction throughput in Proof-of-Work based
Blockchain Networks" was published in DECENTRALIZED 2019 \cite{optimal}, and "UL-blockDAG : Unsupervised Learning based Consensus Protocol for Blockchain" was accepted for publication in The Second International Workshop on Blockchain and Mobile Applications (BlockApp) co-located with 40th IEEE International Conference on Distributed Computing Systems (ICDCS) 2020 \cite{icdcs}.

\bibliography{IEEEabrv,TSC}

\appendices
\section{Proof of Lemma 5.1}
The block creation rate of each miner $j$ follows a \\ poission process
\begin{equation}
\left\lbrace z_j\right\rbrace_{j=1}^{m} \sim
poiss(\lambda_j)
\end{equation}
Hence,
\begin{equation}
\sum_{j=1}^m z_j \sim poiss(\lambda)
\end{equation} 
The block creation interval follows exponential distribution
\begin{align}
T &\sim \exp(\lambda)
\end{align}

The main chain growth rate \cite{ghost} at a height $N$ is
\begin{equation}
\beta \geq \left[\dfrac{1}{N}\sum_{i=1}^{N} X_i \right]^{-1}
\end{equation}

Where, 
\begin{equation}
X_i = time(B_1^{i+1}) - time(B_1^{i})
\end{equation}

Clearly, $X_i$'s are i.i.d random variables because it denotes the time required to add a block to main chain which is equal to $D$ sec for the block to spread plus waiting time  for the creation of next block (randomly distributed as $T$).

\begin{equation}
\left\lbrace X_i\right\rbrace _{i=1}^{N} = D + T
\end{equation}
Thus, 
\begin{align}
\label{eq:meanvarX}
\mu = E\left[ X_i \right] = D + \frac{1}{\lambda},\hspace{0.2cm}
\sigma^2 = VAR\left( X_i \right) = \frac{1}{\lambda^2}
\end{align}

\begin{figure}[t]
\centering
\includegraphics[width=8 cm]{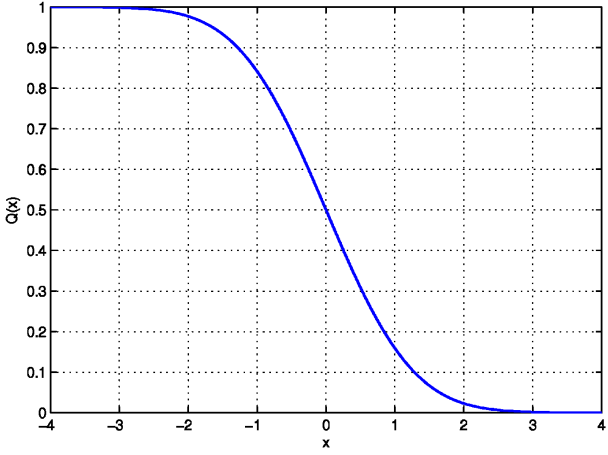}
\caption{The $Q$ function.} 
\label{fig:qfunc}
\end{figure}

Let
\begin{align}
S_N &= \dfrac{1}{N}\sum_{i=1}^{N} X_i 
\end{align}
Using the central limit theorem \cite{chapter8} and \eqref{eq:meanvarX}, 
\begin{align}
S_N \sim \mathcal{N}\left(\mu,\frac{\sigma^2}{N}\right)
\end{align}
Where,
\begin{align}
E\left[S_N\right] &= D + \dfrac{1}{\lambda}= \mu,\hspace{0.2cm} VAR(S_N) = \frac{1}{N\lambda^2} = \frac{\sigma^2}{N}
\end{align}
\begin{align}
\therefore \Pr\Big[\dfrac{1}{S_N} \leq \beta\Big] 
=Q\left(\dfrac{\sqrt{N}}{\sigma}\left( \dfrac{1}{\beta}- \mu\right)\right)
\end{align}
where $Q(\cdot)$ is the $Q$-function \cite{proakis}. From Figure. \ref{fig:qfunc}, it is obvious that the maximum value of $Q(x) \approx 1$ for $x < -(3+\delta)$, $\delta << 1$. \\
Thus, $\Pr\Big[\dfrac{1}{S_N} \leq \beta\Big] $ is maximum for 

\begin{align}
\frac{\sqrt{N}}{\sigma}\left(\frac{1}{\beta}- \mu\right) & \leq -(3+\delta) \\
\implies \beta  \geq \frac{\lambda}{1 - \frac{3+\delta}{\sqrt{N}} + \lambda D }&
\end{align}

%%%%%
\section{Proof of Theorem 5.3}
\begin{figure}[t]
\centering
\includegraphics[width=8 cm]{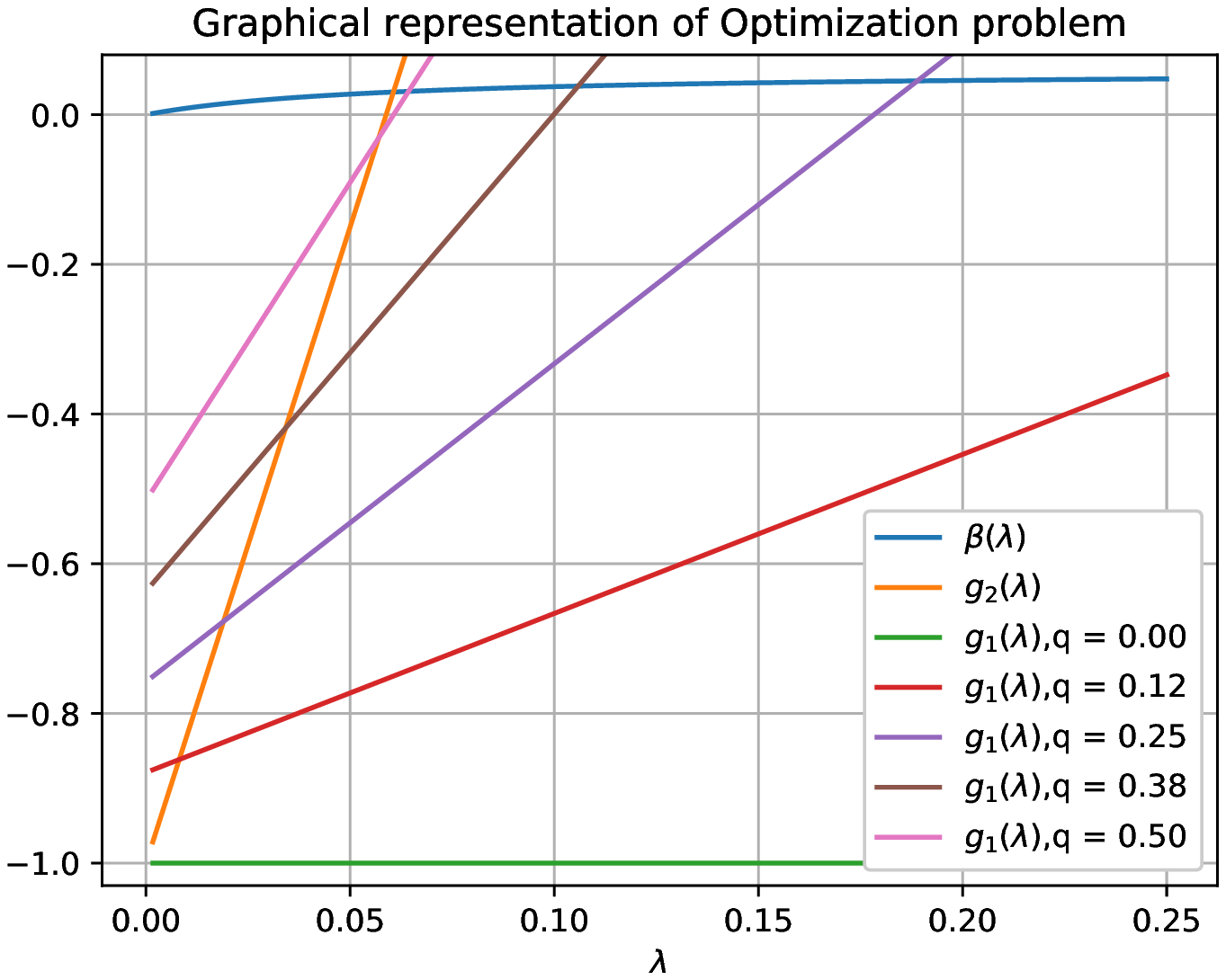}
\caption{} 
\label{fig:opt}
\end{figure}

From the optimization problem , The objective function 
\begin{align}
f(\lambda) = \dfrac{\lambda}{1- \frac{3+\delta}{\sqrt{N}}+\lambda D}
\end{align}
is  concave  and 
and the constraints  
\begin{align}
g_1(\lambda) &= q\left(1- \frac{3+\delta}{\sqrt{N}}+\lambda D\right) -1, \\
g_2(\lambda) &= \lambda D - 1 
\end{align} 
are  affine. \\
The lagrangian \cite{convex} of the  optimization problem is given by 
\begin{equation}
L\left(\lambda,\alpha\right) =  \frac{\lambda}{1- \frac{3+\delta}{\sqrt{N}}+\lambda D} - \mu_1\left(q\left(1- \frac{3+\delta}{\sqrt{N}}+\lambda D\right)-1\right) - 
\mu_2\left(\lambda D - 1\right)
\end{equation}
The optimal solution is obtained by  solving 

\begin{align}
 \dfrac{\partial L\big(\lambda,\alpha\big) }{\partial \lambda}=0 \\
 \implies \dfrac{1- \frac{3+\delta}{\sqrt{N}}}{\Big(1- \frac{3+\delta}{\sqrt{N}}+\lambda D\Big)^2} = \mu_1 q D + \mu_2 D
\label{eq:1}
\end{align}
and
\begin{align}
\dfrac{\partial L\left(\lambda,\alpha\right)}{\partial \mu_1}&=0 
 \\
\implies q\left(1- \frac{3+\delta}{\sqrt{N}}+\lambda D\right) -1 &= 0 
\label{eq:2}
\end{align}
and
\begin{align}
\dfrac{\partial \left(\lambda,\alpha\right)}{\partial \mu_2}&=0  \\
\implies \lambda D-1 &= 0 
\label{eq:3}
\end{align}
From (\ref{eq:2}) and (\ref{eq:3}),
\begin{align}
\label{eq:lamfinal1}
\lambda &= \frac{1}{D} \left(\frac{p}{q}+ \frac{3+\delta}{\sqrt{N}}\right)
\end{align}
\begin{align}
\label{eq:lamfinal2}
\lambda &= \frac{1}{D}
\end{align}
From (\ref{eq:lamfinal1}), 
\begin{align}
g_2(\lambda) > 0
\end{align}
So, $g_1(\lambda) < 0$ is an inactive ($\mu_1 = 0$) and for $q=0$, $g_1(\lambda)$ no longer become a constraint. These can be observed in Fig. \ref{fig:opt}. 

From (\ref{eq:1}) and (\ref{eq:lamfinal2})
\begin{align}
\label{eq:lamfinal}
\lambda = \frac{1}{D},  \quad
\mu_2  = \frac{1-\frac{3+\delta}{\sqrt{N}}}{\left( 2-\frac{3+\delta}{\sqrt{N}}\right)^2}\frac{1}{D}
%\label{eq:mu}
\end{align}
Since $\mu_2 > 0$, after creating sufficiently large number of blocks (N), $\lambda$ in  (\ref{eq:lamfinal}) yields the optimum throughput in Theorem $5.3$ for sufficiently large $N$.

\section{Proof of Lemma 6.1}
The block creation is a Poisson process. Suppose a block $\textbf{B}$ is created at time $t$ with a rate of $\lambda$, during the propagation of the block $\textbf{B}$ to the entire network in a duration of $D$ sec, there are other blocks created in parallel to block $\textbf{B}$ with a rate of $\lambda D$. So, the main chain growth rate $\beta$ at height $N$ is lower bounded by
\begin{equation}
\beta \geq \left[\dfrac{1}{N}\sum_{i=1}^{N} X_i \right]^{-1}
\end{equation}
Where, $X_i$'s are i.i.d random variables which denotes the time interval between the time of creation of first block at $(i+1)^{th}$ height and time of creation of first block at $(i)^{th}$ height
\begin{equation}
X_i = time(B_1^{i+1}) - time(B_1^{i})
\end{equation}
Clearly, this time interval denotes two independent poission processes with rates $\lambda$ and $\lambda D$.
Hence
\begin{equation}
X_i = Y + Z
\end{equation}
Where,
\begin{equation}
Y \sim exp\left(\lambda \right), \hspace*{0.2cm} Z \sim exp\left(\lambda D\right)
\end{equation}
Thus, 
\begin{align}
\label{eq:meanvarT}
\mu &= E\left[X_i \right] = \frac{1}{\lambda} + \frac{1}{\lambda D}, \hspace*{0.1cm}
\sigma^2 = \text{VAR}\left( X_i \right) = \frac{1}{\lambda^2} + \frac{1}{(\lambda D)^2}
\end{align}
Let
\begin{align}
S_N &= \dfrac{1}{N}\sum_{i=1}^{N} X_i 
\end{align}
Using the central limit theorem \cite{chapter8} and \eqref{eq:meanvarT}, 
\begin{align}
S_N \sim \mathcal{N}\left(\mu,\frac{\sigma^2}{N}\right)
\end{align}
Where,
\begin{align}
 E\big[S_N\big] &= \frac{1}{\lambda} + \frac{1}{\lambda D}= \mu,\\ VAR(S_N) &= \frac{1}{N}\left(\frac{1}{\lambda^2} + \frac{1}{(\lambda D)^2}\right) = \frac{\sigma^2}{N}
\end{align}

\begin{align}
%\dfrac{\sqrt{n}}{\sigma}(S_N - \mu) &\sim \mathcal{N}(0,1) \\
\therefore \Pr\Big[\dfrac{1}{S_N} \leq \beta\Big] 
%&= P\Big[S_N > \dfrac{1}{\beta}\Big]\\ &=  P\Big[\dfrac{\sqrt{N}}{\sigma}(S_N - \mu) > \dfrac{\sqrt{N}}{\sigma}\big( \dfrac{1}{\beta}- \mu\big)\Big] \\
=Q\left(\dfrac{\sqrt{N}}{\sigma}\left( \dfrac{1}{\beta}- \mu\right)\right)
\end{align}
where $Q(\cdot)$ is the $Q$-function \cite{proakis}. From Figure. \ref{fig:qfunc}, it is obvious that the maximum value of $Q(x) \approx 1$ for $x < -(3+\delta)$, $\delta << 1$.\\ 
Thus, $\Pr\Big[\dfrac{1}{S_N} \leq \beta\Big] $ is maximum for 
\begin{align}
\frac{\sqrt{N}}{\sigma}\left(\frac{1}{\beta}- \mu\right) & \leq -(3+\delta) \\
\implies \beta  &\geq \frac{\lambda D}{1 + D - \frac{(3+\delta)D }{\sqrt{N(1+D^2)}}} \geq \frac{\lambda D}{1 + D - \frac{(3+\delta)}{\sqrt{N}}}
\end{align}
\section{Proof of Lemma 6.4}
Suppose a block $\textbf{C}$ is the first reference to block $\textbf{B}$ by any honest node at height $N+1$, during the propagation of the block $\textbf{C}$ to the entire network in a time $D$, there are other honest blocks (that refers to $\mathbf{B}$) created in parallel to block $\textbf{C}$ with a rate of $\lambda (1-q) D$. So, block references to block $\textbf{B}$ is a uniform random variable given Poission random variable with a rate $r_1 = \lambda(1-q)D$.  

Let 
\begin{equation}
P \sim Poiss(r_1), \quad R_{h,i}\big|P  \sim U\left[1,P\right]
\end{equation}

As per \textbf{Definition 7}, $R_{a,N+k+1}$ represents the number of blocks in \textit{future}($\textbf{B}^{'}$) at a height $N+k+1$. Since $R_{a,N+k+1}$ consists of the blocks from both honest blockDAG and attacker's secret chain, $R_{a,N+k+1}$ is also a uniform random variable given Poisson random variable with a rate $r_2 = \lambda D$.

Let
\begin{equation}
P' \sim Poiss(r_2), \quad R_{a,N+k+1}\big|P' \sim U\left[1,P'\right]
\end{equation}
The Expectation and variance of these random variables using the law of iterated expectation and law of total variance \cite{prob} are given in Table \ref{table:E-V}.
\begin{table}[!t]
\caption{\\ Expectation and Variance}
  \label{table:E-V}
\centering
\begin{tabular}{|c |l |l |}
\hline
    \textbf{Random Variable}&\textbf{E[.]}&VAR(.)\\
\hline
    $P$ & $ r_1$ & $r_1$    \\
    \hline
    $R_{h,i}\big|P$ & $(P+1)/2$ & $(P^2 - 1)/12$\\
    \hline
    $R_{h,i}$ & $(r_1+1)/2$ & $((r_1 + 1)^2 + 2(r_1 - 1))/12 $ \\
    \hline
    $P'$  & $r_2$ &  $r_2$ \\
    \hline
    $R_{a,N+k+1}\big|P'$ & $(P'+1)/2$ & $(P'^2 - 1)/12$\\
    \hline
    $R_{a,N+k+1}$ & $(r_2 + 1)/2$ & $((r_2 + 1)^2 + 2(r_2 - 1))/12$\\
    	\hline
		  \end{tabular}
\end{table}

For a $Risk(\mathbf{B},k)$ of atmost $\epsilon$,
\begin{align}
Pr\left[\sum_{i=N+1}^{N+k+1}R_{h,i} < R_{a,N+k+1}\right] &< \epsilon \\
Pr\left[\sum_{i=N+1}^{N+k+1}R_{h,i} \geq R_{a,N+k+1} \right] &\geq 1- \epsilon
\end{align}
Let 
$X = \sum_{i=N+1}^{N+k+1}R_{h,i}$, $Y = R_{a,N+k+1}$, $f(X,Y) = \frac{X}{Y}$, then
\begin{align}
Pr\left[\frac{X}{Y} < 1 \right] \geq 1-\epsilon
\label{eq:ineq} 
\end{align}
Using the taylor series approaximation of $f(X,Y)$ \cite{ratio-expectation},
\begin{align}
E\left[f(X,Y)\right] &\approx \frac{k(r_1 + 1)}{r_2 + 1} + \frac{(r_2 + 1)^2 + 2(r_2 -1)}{12} \frac{4k(r_1 + 1)}{(r_2 + 1)^3} \\
&\approx \frac{k(r_1 + 1)}{r_2 + 1} + \frac{k(r_1 + 1)}{3(r_2 + 1)} + \frac{2k(r_2 - 1)(r_1 + 1))}{3(r_2 +1)^3}
\end{align}
Since $\lambda D >> 1$ for UL-blockDAG protocol, 
\begin{align}
E\left[f(X,Y)\right] \approx \frac{4k(r_1 + 1)}{3(r2+1)} = \frac{4k\left(\lambda (1-q)D + 1\right)}{3(\lambda D+1)}
\label{eq:E} 
\end{align}
Using Markov's inequality and \eqref{eq:E}
\begin{align}
k \geq \frac{3(\lambda D +1)(1-\epsilon)}{4\left[\lambda (1-q)D + 1\right]}
\end{align}

\vskip 0\baselineskip plus -1fil
\begin{IEEEbiography}[{\includegraphics[width=1.25in,height=1.05in,clip,keepaspectratio]{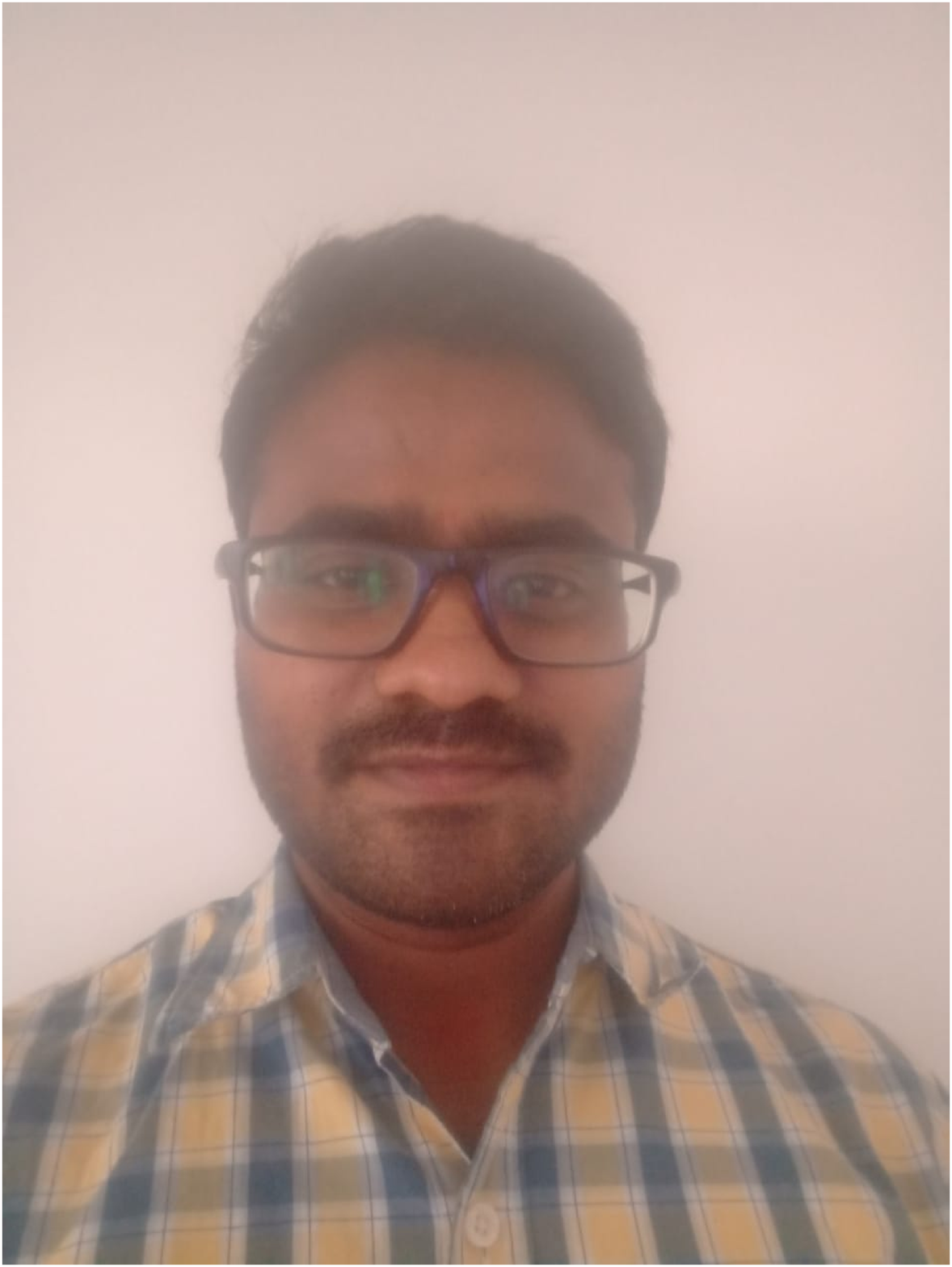}}]{B Swaroopa reddy}
is currently pursuing the Ph.D degree in the Department of Electrical Engineering, Indian Institute of Technology Hyderabad, India from 2017. He received the B.Tech degree in Electronics and Communication Engineering from Sri Krishna Devaraya University, Anantapur, India, in 2009. From 2010 to 2017, he was with the Bharath Sanchar Nigam Limited (BSNL), a Public Sector Unit under the Government of India. His research interests include scalability and security in Blockchain networks and Integration of IoT with Blockchain. He is a member of the IEEE. 
\end{IEEEbiography}
\vskip -1\baselineskip plus -1fil
\begin{IEEEbiography}[{\includegraphics[width=1in,height=1.25in,clip,keepaspectratio]{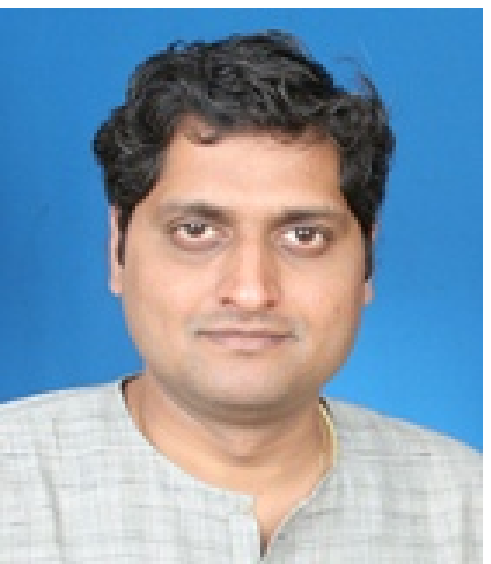}}]{GVV Sharma}is an Associate Professor in Department of Electrical Engineering, Indian Institute of Technology, Hyderabad, India. He received the PhD degree in Electrical Engineering Department from Indian Institute of Technology Bombay, India, in 2010, the M.Sc. (Eng.) degree in electrical communication engineering from the Indian Institute of Science, Bangalore, India, in 2004 and the B.Tech. degree in Electronics and communication engineering from the Indian Institute of Technology, Guwahati, India, in 1999. From August 2004 to July 2006, he was with the Applied Research Group of Satyam Computers, Bangalore, India. His interests  are in Developmental Engineering. He is a member of the IEEE.
\end{IEEEbiography}

\end{document}